\documentclass[11pt]{article}
\usepackage{amsmath,amssymb,amsthm,epic,eepic,graphicx}
\usepackage{subfigure}

\advance\textwidth 20mm  \advance\oddsidemargin -10mm
\advance\textheight 20mm \advance\topmargin -10mm

\theoremstyle{plain}

\newtheorem{theorem}{Theorem}

\theoremstyle{definition}
\newtheorem{definition}{Definition}

\theoremstyle{remark}

\def\setcircle#1{\def\rrr{#1}}
\def\Black(#1,#2){\put(#1,#2){\circle*{\rrr}}}
\def\White(#1,#2){\put(#1,#2){\circle{\rrr}}}

\newcommand{\wx}{\widetilde{x}}

\newcommand{\eto}[1]{e^{\displaystyle #1}}
\newcommand{\nm}{\!-\!}

\newcommand{\ueto}[1]{e^{\raisebox{2pt}{$\displaystyle #1$}}}

\newbox\meibox
\def\placeunder#1#2#3#4{\setbox\meibox%
\vbox{\hbox{\hskip#4$\hphantom{#2}$}\hbox{$\hphantom{#1}$}}%
\vtop{\baselineskip=0pt\lineskiplimit=\baselineskip%
\lineskip=#3\hbox to \wd\meibox{\hfil\hskip#4$#2$\hfil}%
\hbox to \wd\meibox{\hfil$#1$\hfil}}}
\def\undertilde#1{\mathchoice{%
\placeunder{\vbox to 1.4pt{\hbox{$\displaystyle\widetilde{\,\,\,
}$}\vss}}{\displaystyle#1}{1.5pt}{1.5pt}}%
{\placeunder{\vbox to 1.4pt{\hbox{$\textstyle\widetilde{\,\,
}$}\vss}}{\textstyle#1}{1.5pt}{1.5pt}}%
{\placeunder{\vbox to 1.4pt{\hbox{$\scriptstyle\tilde{
}$}\vss}}{\scriptstyle#1}{1pt}{1pt}}%
{\placeunder{\vbox to 1.4pt{\hbox{$\scriptscriptstyle\tilde{
}$}\vss}}{\scriptscriptstyle#1}{1pt}{1pt}}%
}

\def\cG{{\cal G}}
\def\cD{{\cal D}}
\def\cT{{\cal T}}
\def\cK{{\cal K}}

\def\fe{{\mathfrak e}}

\begin{document}

\markboth{R. Boll and Yu.B. Suris}{Discrete Toda systems}

\title{Non-symmetric discrete Toda systems from quad-graphs}

\author{Raphael Boll$^{1}$ and Yuri B. Suris$^{2}$}

\maketitle

\footnotetext[1]{Zentrum Mathematik, Technische Universit\"at
M\"unchen, Boltzmannstr. 3, 85748 Garching, Germany}
\footnotetext[2]{Institut f\"{u}r Mathematik, MA 7-2, Technische
Universit\"at Berlin, Str. des 17. Juni 136, 10623 Berlin,
Germany}

\begin{abstract}
For all non-symmetric discrete relativistic Toda type equations we
establish a relation to 3D consistent systems of quad-equations.
Unlike the more simple and better understood symmetric case, here
the three coordinate planes of $\mathbb Z^3$ carry different
equations. Our construction allows for an algorithmic derivation
of the zero curvature representations and yields analogous results
also for the continuous time case.
\end{abstract}

\section{Introduction}
\label{Sect intro}

This paper is devoted to one aspect of the general topic of
discrete integrable systems. In the recent years the viewpoint
that discrete integrable systems are in a sense more fundamental
than continuous ones becomes gradually more and more accepted in
the soliton theory community. The last developments led to the
understanding that the very definition of integrability becomes
more transparent and natural on the discrete level, see
\cite{BS08}. One can consider certain types of discrete equations,
like the so called quad-equations, as a sort of ``elementary
particles'' of the solitonic world, with the enormous richness of
this world resulting from various combinations and limiting
procedures of these elementary objects. Here, we discover such
``elementary particles'' underlying the so called equations of the
relativistic Toda type. This ``microstructure'' become visible
only after the discretization procedure, and even then some hidden
degrees of freedom remain to be uncovered in order for the whole
simplicity to become apparent.

We review the relativistic Toda type equations and their
integrable discretizations in Sections \ref{Sect RTL} and
\ref{Sect dRTL}, respectively. After that, general Toda equations
on graphs are discussed in Section \ref{Sect: dToda}. Fundamental
notion of quad-graphs and quad-equations (which belong, in our
view, to the ``elementary particles'' mentioned above) are
discussed in Section \ref{Sect quad-graphs}, while the following
Section \ref{Sect: Q to Toda} is devoted to the explanation of how
the Toda type equations on graphs can be reduced to the systems of
quad-equations. In Section \ref{s:7TL} some combinatorial aspects
of the regular triangular lattice are discussed, since this
lattice underlies the discrete relativistic Toda type equations.
Finally, Sections \ref{Sect main}--\ref{Sect zcr} contain the main
new results of this paper. Namely, in Section \ref{Sect main} we
present the systems of quad-equations which yield the
non-symmetric discrete relativistic Toda equations. In Section
\ref{Sect master} we show that all this systems have their common
origin in just one master system of this type. In Section
\ref{Sect zcr} we demonstrate how to derive the zero curvature
representations for relativistic Toda type equations (discrete and
continuous) in an algorithmic manner. A brief outlook is
formulated in the concluding Section \ref{Sect concl}.

\section{Lattice equations of the relativistic Toda type}
\label{Sect RTL}

The term ``equations of the relativistic Toda type'' is used to
denote integrable lattice equations of the general form
\begin{eqnarray}\label{RTL gen intro}
\ddot{x}_k & = & r(\dot{x}_k)\Big(\dot{x}_{k+1}f(x_{k+1}-x_k)-
\dot{x}_{k-1}f(x_k-x_{k-1})\nonumber\\
&& \qquad\quad+g(x_{k+1}-x_k)-g(x_k-x_{k-1})\Big).
\end{eqnarray}
The relativistic Toda lattice proper was invented by
S.~Ruijsenaars \cite{Ru90}. It is described by Newtonian equations
of motion
\begin{eqnarray}\label{RTL+ lq New introd}
\lefteqn{\ddot{x}_k=
(1+\alpha\dot{x}_{k+1})(1+\alpha\dot{x}_k)\frac{\eto{x_{k+1}\nm
x_k}} {\,\raisebox{-1mm}{$1+\alpha^2\eto{x_{k+1}\nm
x_k}$}}}\nonumber\\
&&\qquad\qquad-(1+\alpha\dot{x}_k)(1+\alpha\dot{x}_{k-1})\frac{\eto{x_k\nm
x_{k-1}}} {\,\raisebox{-1mm}{$1+\alpha^2\eto{x_k\nm x_{k-1}}$}}\;.
\qquad
\end{eqnarray}
Here $\alpha$ is a small parameter whose physical meaning is the
inverse speed of light. In the non-relativistic limit $\alpha\to
0$ system (\ref{RTL+ lq New introd}) turns into the usual Toda
lattice.

It took about a decade for further integrable equations of the
relativistic Toda type to be discovered. In \cite{Su97} the
following ones were found: two systems which again can be
considered as $\alpha$-perturbations of the usual Toda lattice:
\begin{eqnarray}\label{RTL+ l New introd}
\ddot{x}_k & = & (1+\alpha\dot{x}_{k+1})\,\eto{x_{k+1}\nm x_k}-
(1+\alpha\dot{x}_{k-1})\,\eto{x_k\nm x_{k-1}}\nonumber\\
 & & -\alpha^2\eto{2(x_{k+1}\nm x_k)}+\alpha^2\eto{2(x_k\nm x_{k-1})}
\end{eqnarray}
and
\begin{equation}\label{RTL- l New introd}
\ddot{x}_k=(1-\alpha\dot{x}_k)^2\Big((1-\alpha\dot{x}_{k+1})\,
\eto{x_{k+1}\nm x_k}-(1-\alpha\dot{x}_{k-1})\,\eto{x_k\nm
x_{k-1}}\Big),
\end{equation}
and two systems which can be considered as $\alpha$-perturbations
of the so called modified Toda lattice:
\begin{eqnarray}\label{RTL+ m New introd}
\ddot{x}_k & = & \dot{x}_k\Big(\eto{x_{k+1}\nm x_k}-\eto{x_k\nm
x_{k-1}}\Big)
\nonumber\\
 & & +\alpha\left(\dot{x}_{k+1}\dot{x}_k\,\frac{\eto{x_{k+1}\nm x_k}}
{\,\raisebox{-1mm}{$1+\alpha\eto{x_{k+1}\nm x_k}$}}
-\dot{x}_k\dot{x}_{k-1}\,\frac{\eto{x_{k}\nm x_{k-1}}}
{\,\raisebox{-1mm}{$1+\alpha\eto{x_{k}\nm
x_{k-1}}$}}\right)\qquad\qquad
\end{eqnarray}
and
\begin{equation}\label{RTL- m New introd}
\ddot{x}_k =\dot{x}_k(1-\alpha\dot{x}_k)\!\left(\!
(1-\alpha\dot{x}_{k+1})\,\frac{\eto{x_{k+1}\nm x_k}}
{\raisebox{-1mm}{$1+\alpha\eto{x_{k+1}\nm x_k}$}}-
(1-\alpha\dot{x}_{k-1})\,\frac{\eto{x_k\nm x_{k-1}}}
{\raisebox{-1mm}{$1+\alpha\eto{x_k\nm x_{k-1}}$}}\!\right).
\end{equation}
For all these systems a Lagrangian and a Hamiltonian formulations
were given, and their complete integrability was demonstrated by
presenting the full set of integrals of motion and a local zero
curvature representations of the type
\begin{equation}\label{zcr intro}
\dot{L}_k=M_{k+1}L_k-L_kM_k
\end{equation}
in terms of $2\times 2$ matrices (locality means that the matrix
$L_k$ depends only on $x_k$ and the corresponding canonically
conjugate momentum $p_k$, and not on phase variables from other
lattice sites). Two further systems of the relativistic Toda type
with rational interactions (as opposed to exponential interactions
in the previous ones) appeared in \cite{Su99}:
\begin{equation}\label{RTL+ dual New introd}
\ddot{x}_k=\dot{x}_k\left(x_{k+1}-2x_k+x_{k-1}\right)+
\frac{\alpha\dot{x}_{k+1}\dot{x}_k}{1+\alpha(x_{k+1}-x_k)}-
\frac{\alpha\dot{x}_k\dot{x}_{k-1}}{1+\alpha(x_k-x_{k-1})}
\end{equation}
and
\begin{equation}\label{RTL- dual New introd}
\ddot{x}_k=\dot{x}_k(1+\alpha^2\dot{x}_k)
\left(\frac{x_{k+1}-x_k-\alpha\dot{x}_{k+1}}{1+\alpha(x_{k+1}-x_k)}-
\frac{x_k-x_{k-1}-\alpha\dot{x}_{k-1}}{1+\alpha(x_k-x_{k-1})}\right).
\end{equation}
Both these systems are $\alpha$-perturbations of the so-called
dual Toda lattice. Also for these systems the Lagrangian and the
Hamiltonian formulations were given in \cite{Su99}, as well as a
demonstration of their complete integrability. However, local zero
curvature representations were not given at that point.

Along another line of research, a complete classification of
``integrable'' systems of the type (\ref{RTL gen intro}) was
achieved in \cite{ASh97}. The notion of ``integrability'' used in
this paper is a clever and unexpected device, allowing to carry
out a complete classification, but it has, {\it \`a priori},
nothing to do with the usual Liouville-Arnold integrability.
Namely, they noticed that the above systems are always Lagrangian,
and required that their form be retained under a sort of Legendre
transformation. This allowed them to find all the Newtonian
equations mentioned above, as well a new series of systems,
including the most general one,
\begin{eqnarray}\label{RTL cq New introd}
\ddot{x}_k & = & -\frac{1}{2}\,(\dot{x}_k^2-\nu^2)
\left(\frac{\sinh\, 2(x_{k+1}-x_k)-
\nu^{-1}\sinh(2\nu\alpha)\,\dot{x}_{k+1}}
{\raisebox{-1mm}{$\sinh^2(x_{k+1}-x_k)-\sinh^2(\nu\alpha)$}}\;- \right. \nonumber\\
 & & \qquad-\left.\frac{\sinh\, 2(x_k-x_{k-1})-
\nu^{-1}\sinh(2\nu\alpha)\,\dot{x}_{k-1}}
{\raisebox{-1mm}{$\sinh^2(x_k-x_{k-1})-\sinh^2(\nu\alpha)$}}\right),
\end{eqnarray}
its limiting case (first rescale $x_k\mapsto\nu x_k$,
$\nu\mapsto\gamma\nu$, and then send $\nu\to 0$):
\begin{equation}\label{RTL c2 New introd}
\ddot{x}_k=-(\dot{x}_k^2-\gamma^2)\left(
\frac{x_{k+1}-x_k-\alpha\dot{x}_{k+1}}{(x_{k+1}-x_k)^2-\gamma^2\alpha^2}
-\frac{x_k-x_{k-1}-\alpha\dot{x}_{k-1}}{(x_k-x_{k-1})^2-\gamma^2\alpha^2}
\right),
\end{equation}
as well as the particular cases $\nu=0$, resp. $\gamma=0$, of the
latter two systems:
\begin{eqnarray}\label{RTL c New introd}
\ddot{x}_k & = &
-\dot{x}_k^2\Bigg(\coth(x_{k+1}-x_k)-\coth(x_k-x_{k-1}) \nonumber\\
& & \qquad
    -\,\frac{\alpha\dot{x}_{k+1}}{\raisebox{-1mm}{$\sinh^2(x_{k+1}-x_k)$}}
    +\frac{\alpha\dot{x}_{k-1}}{\raisebox{-1mm}{$\sinh^2(x_k-x_{k-1})$}}\Bigg)
\end{eqnarray}
and
\begin{equation}\label{RTL c3 New introd}
\ddot{x}_k=-\dot{x}_k^2
\left(\frac{1}{x_{k+1}-x_k}-\frac{1}{x_k-x_{k-1}}-
\frac{\alpha\dot{x}_{k+1}}{(x_{k+1}-x_k)^2}+
\frac{\alpha\dot{x}_{k-1}}{(x_k-x_{k-1})^2}\right).
\end{equation}
Complete integrability of these systems in the usual sense, along
with $2\times 2$ local zero curvature representations for the last
two ones, was demonstrated in the monograph \cite{Su03}.

\section{Time discretization of the relativistic Toda type equations}
\label{Sect dRTL}

Integrable discretizations of the relativistic Toda type equations
all have the following general shape of {\it discrete time
Newtonian equations of motion}:
\begin{eqnarray}\label{dRTL gen intro}
\lefteqn{F(\wx_k-x_k)-F(x_k-\undertilde{x}_k)=}\nonumber\\
&& G(x_{k+1}-x_k)-G(x_k-x_{k-1})+
H(\undertilde{x}_{k+1}-x_k)-H(x_k-\wx_{k-1}).
\end{eqnarray}
Here and below we use the following abbreviations for functions of
the discrete time $h\mathbb Z$:
\[
x_k=x_k(t),\quad \wx_k=x_k(t+h),\quad \undertilde{x}_k=x_k(t-h).
\]
The integrability preserving time discretization of the
Ruijsenaars' relativistic Toda lattice was performed in
\cite{Su96}, where the following equations were derived:
\begin{equation}\label{dRTL+ lq New introd}
\displaystyle \frac{1+\alpha h^{-1}\bigg(\eto{\wx_k\nm
x_k}-1\bigg)} {1+\alpha h^{-1}\bigg(\ueto{x_k\nm
\undertilde{x}_k}-1\bigg)}=
\displaystyle\frac{\bigg(1+\alpha^2\eto{x_{k+1}\nm x_k}\bigg)
\bigg(1+\alpha(\alpha-h)\eto{x_k\nm\wx_{k-1}}\bigg)}
{\bigg(1+\alpha^2\ueto{x_k\nm x_{k-1}}\bigg)
\bigg(1+\alpha(\alpha-h)\ueto{\undertilde{x}_{k+1}-x_k}\bigg)}\;.
\end{equation}
It was shown in \cite{Su96} that, upon a natural Lagrangian (or
discrete Hamiltonian) re-formulation in terms of the canonically
conjugate variables $x_k,p_k$, discrete time equations (\ref{dRTL+
lq New introd}) share integrals of motion with the continuous time
equations (\ref{RTL+ lq New introd}), and therefore have the same
integrability properties (belong to the same integrable
hierarchy). This remains true also for all integrable
discretizations in this section.

The Newtonian systems (\ref{RTL+ l New introd}) and (\ref{RTL- l
New introd}) were discretized in \cite{Su97} in an ``additive''
manner as
\begin{eqnarray}\label{dRTL+ l New introd}
\ueto{\wx_k\nm x_k}-\ueto{x_k\nm\undertilde{x}_k} & = &
h\alpha\ueto{x_{k+1}\nm x_k}-h\alpha\eto{x_k\nm x_{k-1}}
\nonumber\\
 & &-\displaystyle\frac{h(\alpha-h)\,\ueto{\undertilde{x}_{k+1}-x_k}}
{\,\raisebox{-2mm}{$1-h\alpha\ueto{\undertilde{x}_{k+1}-x_k}$}}
+\displaystyle\frac{h(\alpha-h)\,\eto{x_k\nm \wx_{k-1}}}
{\,\raisebox{-2mm}{$1-h\alpha\eto{x_k\nm \wx_{k-1}}$}}\qquad\quad
\end{eqnarray}
and
\begin{eqnarray}\label{dRTL- l New introd}
\frac{1}{1-\alpha h^{-1}\bigg(\eto{\wx_k\nm x_k}-1\bigg)}\,-\,
\frac{1}{1-\alpha
h^{-1}\bigg(\ueto{x_k\nm\undertilde{x}_k}-1\bigg)}=
\qquad\qquad\qquad \nonumber\\
=\alpha(\alpha+h)\ueto{\undertilde{x}_{k+1}-x_k}
-\alpha(\alpha+h)\ueto{x_k\nm \wx_{k-1}} -\alpha^2\ueto{x_{k+1}\nm
x_k}+\alpha^2\ueto{x_k\nm x_{k-1}},\qquad
\end{eqnarray}
while the discretizations of the systems (\ref{RTL+ m New introd})
and (\ref{RTL- m New introd}) given there were ``multiplicative'':
\begin{equation}\label{dRTL+ m New introd}
\displaystyle\frac{\bigg(\ueto{\wx_k\nm x_k}-1\bigg)}
{\bigg(\ueto{x_k\nm\undertilde{x}_k}-1\bigg)} =
\displaystyle\frac{\bigg(1+h\ueto{\undertilde{x}_{k+1}-x_k}\bigg)
\bigg(1+\alpha\eto{x_k\nm\wx_{k-1}}\bigg)
\bigg(1+\alpha\eto{x_{k+1}\nm x_k}\bigg)}
{\bigg(1+\alpha\ueto{\undertilde{x}_{k+1}-x_k}\bigg)
\bigg(1+h\eto{x_{k}\nm\wx_{k-1}}\bigg) \bigg(1+\alpha\eto{x_k\nm
x_{k-1}}\bigg)}
\end{equation}
and
\begin{eqnarray}\label{dRTL- m New introd}
\lefteqn{\frac{\bigg(\eto{\wx_k\nm x_k}-1\bigg)}
{\bigg(\ueto{x_k\nm \undertilde{x}_k}-1\bigg)}\cdot \frac{1-\alpha
h^{-1}\bigg(\ueto{x_k\nm \undertilde{x}_k}-1\bigg)} {1-\alpha
h^{-1}\bigg(\eto{\wx_k\nm x_k}-1\bigg)}=}\nonumber\\
&&=\displaystyle\frac{\bigg(1+\alpha\ueto{x_k\nm x_{k-1}}\bigg)}
{\bigg(1+\alpha\eto{x_{k+1}\nm x_k}\bigg)} \cdot\displaystyle\frac
{\bigg(1+(\alpha+h)\,\ueto{\undertilde{x}_{k+1}-x_k}\bigg)}
{\bigg(1+(\alpha+h)\,\eto{x_k\nm\wx_{k-1}}\bigg)}\;.
\end{eqnarray}
Discretizations of the rational systems (\ref{RTL+ dual New
introd}) and (\ref{RTL- dual New introd}) appeared in \cite{Su99}:
\begin{equation}\label{dRTL+ dual New introd}
\frac{\wx_k-x_k}{x_k-\undertilde{x}_k}=
\frac{\Big(1+h(\undertilde{x}_{k+1}-x_k)\Big)}
     {\Big(1+\alpha(\undertilde{x}_{k+1}-x_k)\Big)}\cdot
\frac{\Big(1+\alpha(x_k-\wx_{k-1})\Big)}
     {\Big(1+h(x_k-\wx_{k-1})\Big)}\cdot
\frac{\Big(1+\alpha(x_{k+1}-x_k)\Big)}
     {\Big(1+\alpha(x_k-x_{k-1})\Big)}\quad
\end{equation}
and
\begin{eqnarray}\label{dRTL- dual New introd}
\frac{\,\raisebox{1.5mm}{$(\wx_k-x_k)\,\Big(1+\alpha(\alpha+h)h^{-1}
(x_k-\undertilde{x}_k)\Big)$}}
{\,\raisebox{-1mm}{$(x_k-\undertilde{x}_k)\,\Big(1+\alpha(\alpha+h)h^{-1}
(\wx_k-x_k)\Big)$}}=\qquad\qquad\qquad\qquad\qquad\qquad
\nonumber\\
=\frac{\,\raisebox{1mm}{$\Big(1+\alpha(x_k-x_{k-1})\Big)$}}
{\,\raisebox{-1mm}{$\Big(1+\alpha(x_{k+1}-x_k)\Big)$}}\cdot
\frac{\,\raisebox{1mm}{$\Big(1+(\alpha+h)(\undertilde{x}_{k+1}-x_k)\Big)$}}
{\,\raisebox{-1mm}{$\Big(1+(\alpha+h)(x_k-\wx_{k-1})\Big)$}}\;.\qquad
\end{eqnarray}

A discrete version of the device from \cite{ASh97} was developed
in \cite{A99}. It was used to classify ``integrable'' systems of
the general type (\ref{dRTL gen intro}), i.e., those retaining
their form under a sort of a discrete Legendre transformation. The
resulting list consisted essentially of the systems quoted in
above in this section, as well as of discretizations of the
systems (\ref{RTL cq New introd})--(\ref{RTL c3 New introd}). A
discretization of (\ref{RTL cq New introd}) reads
\begin{eqnarray}\label{dRTL cq New introd}
\lefteqn{\frac{\raisebox{1mm}{$\sinh(\wx_k-x_k+\nu h)$}}
{\raisebox{-1mm}{$\sinh(\wx_k-x_k-\nu h)$}}\cdot
\frac{\raisebox{1mm}{$\sinh(x_k-\undertilde{x}_k-\nu h)$}}
{\raisebox{-1mm}{$\sinh(x_k-\undertilde{x}_k+\nu h)$}}}
\nonumber\\\nonumber\\ &=&
\frac{\raisebox{1mm}{$\sinh(x_{k+1}-x_k+\nu\alpha)$}}
{\raisebox{-1mm}{$\sinh(x_{k+1}-x_k-\nu\alpha)$}}
\cdot\frac{\raisebox{1mm}{$\sinh(x_k-x_{k-1}-\nu\alpha)$}}
{\raisebox{-1mm}{$\sinh(x_k-x_{k-1}+\nu\alpha)$}}\,\nonumber\\\nonumber\\
 && \times
\frac{\raisebox{1mm}{$\sinh(\undertilde{x}_{k+1}-x_k-\nu\alpha+\nu
h)$}}
{\raisebox{-1mm}{$\sinh(\undertilde{x}_{k+1}-x_k+\nu\alpha-\nu
h)$}} \cdot\frac{\raisebox{1mm}{$\sinh(x_k-\wx_{k-1}+\nu\alpha-\nu
h)$}} {\raisebox{-1mm}{$\sinh(x_k-\wx_{k-1}-\nu\alpha+\nu
h)$}}\;.\qquad
\end{eqnarray}
Its rational version, which is a discretization of (\ref{RTL c2
New introd}), reads
\begin{eqnarray}\label{dRTL c2 New introd}
\frac{\raisebox{1mm}{$(\wx_k-x_k+\gamma h)$}}
{\raisebox{-1mm}{$(\wx_k-x_k-\gamma h)$}}\cdot
\frac{\raisebox{1mm}{$(x_k-\undertilde{x}_k-\gamma h)$}}
{\raisebox{-1mm}{$(x_k-\undertilde{x}_k+\gamma h)$}}  =
\frac{\raisebox{1mm}{$(x_{k+1}-x_k+\gamma\alpha)$}}
{\raisebox{-1mm}{$(x_{k+1}-x_k-\gamma\alpha)$}}
\cdot\frac{\raisebox{1mm}{$(x_k-x_{k-1}-\gamma\alpha)$}}
{\raisebox{-1mm}{$(x_k-x_{k-1}+\gamma\alpha)$}}\,\times\nonumber
\\ \nonumber\\
\times\,
\frac{\raisebox{1mm}{$(\undertilde{x}_{k+1}-x_k-\gamma\alpha+\gamma
h)$}}
{\raisebox{-1mm}{$(\undertilde{x}_{k+1}-x_k+\gamma\alpha-\gamma
h)$}}
\cdot\frac{\raisebox{1mm}{$(x_k-\wx_{k-1}+\gamma\alpha-\gamma
h)$}} {\raisebox{-1mm}{$(x_k-\wx_{k-1}-\gamma\alpha+\gamma
h)$}}\;.\qquad
\end{eqnarray}
Finally, the $\nu\to 0$, resp. the $\gamma\to 0$ limits of the
latter two equations lead to additive ones:
\begin{eqnarray}\label{dRTL c New introd}
h\,\coth(\wx_k-x_k)-h\,\coth(x_k-\undertilde{x}_k)=
\alpha\,\coth(x_{k+1}-x_k)-\alpha\,\coth(x_k-x_{k-1})- \nonumber\\
-(\alpha-h)\,\coth(\undertilde{x}_{k+1}-x_k)+
(\alpha-h)\,\coth(x_k-\wx_{k-1})\qquad\qquad
\end{eqnarray}
and
\begin{equation}\label{dRTL c3 New introd}
\frac{h}{\wx_k-x_k}-\frac{h}{x_k-\undertilde{x}_k}=
\frac{\alpha}{x_{k+1}-x_k}-\frac{\alpha}{x_k-x_{k-1}}
-\frac{\alpha-h}{\undertilde{x}_{k+1}-x_k}+\frac{\alpha-h}{x_k-\wx_{k-1}}\;,
\end{equation}
which discretize (\ref{RTL c New introd}) and (\ref{RTL c3 New
introd}), respectively. The integrability of the difference
equations (\ref{dRTL cq New introd})--(\ref{dRTL c3 New introd})
in the usual sense was dealt with in \cite{Su03}, where it was
shown that they share the integrals of motion and the matrices
$L_k$ from the zero curvature representations with their
continuous time counterparts.

\section{Discrete Toda type equations on graphs}
\label{Sect: dToda}

\begin{figure}[t]
\begin{center}
\setlength{\unitlength}{0.035em}
\begin{picture}(450,270)(-25,-30)
 \path(-25,200)(325,200)
 \multiput(0,0)(100,0){4}{
  \Black(0,200)\path(0,100)(0,200)(100,100)
  \Black(100,0)\path(0,100)(100,0)(100,100)}
 \path(-25,100)(425,100)
 \path(-25,125)(0,100)
 \path(400,100)(400,125)
 \multiput(0,100)(100,0){5}{\Black(0,0)}
 \path(0,75)(0,100) \path(75,0)(425,0) \path(400,100)(425,75)
 \multiputlist(100,210)(100,0)[cb]{$\wx_{k-1}$,$\wx_k$}
 \multiputlist(110,110)(100,0)[lb]{$x_{k-1}$,$x_k$,$x_{k+1}$}
 \multiputlist(200,-10)(100,0)[ct]{$\undertilde{x}_k$,$\undertilde{x}_{k+1}$}
\end{picture}
\caption{Regular triangular lattice underlying discrete
relativistic Toda type systems} \label{fig: tl}
\end{center}
\end{figure}
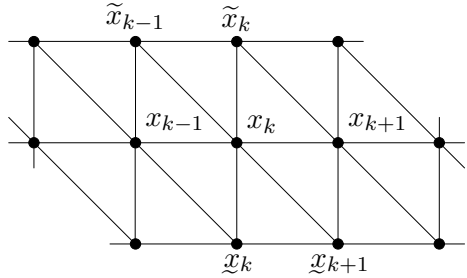

An important observation made in \cite{A00} was that the natural
combinatorial structure underlying the discrete relativistic Toda
type lattices (\ref{dRTL gen intro}) is actually the regular
triangular lattice in the plane (rather than the standard square
lattice $\mathbb Z^2$). Namely, each equation of the system
(\ref{dRTL gen intro}) relates seven fields assigned to the star
of a vertex of the regular triangular lattice, each one of the
functions $F,G,H$ being associated to edges of one of the tree
directions, see Figure \ref{fig: tl}.
%-----------------------------------------------------------------
\setcircle{10}

Note that while in the systems (\ref{dRTL+ lq New
introd})--(\ref{dRTL- dual New introd}) (called hereafter {\em
non-symmetric discrete relativistic Toda type equations}) the tree
functions $F,G,H$ are, generally speaking, different, this is no
more the case for systems (\ref{dRTL cq New introd})--(\ref{dRTL
c3 New introd}). In the latter systems (called hereafter {\em
symmetric discrete relativistic Toda type equations}) all three
functions essentially coincide, differing only by the values of
the built-in parameters.

This allows, at least for symmetric systems, the next
generalization step, which was made in \cite{A01}, namely the
introduction of discrete Toda type systems on arbitrary graphs.

\begin{definition}\label{Dfn:dLaplace}
Let $\cG$ be a graph, with the set of vertices $V(\cG)$ and the
set of edges $E(\cG)$. A {\em discrete Toda type system\,} on
$\cG$ for a function $x:V(\cG)\to\mathbb C$ reads:
\begin{equation}\label{eq:nlin Laplace}
\sum_{v\in\,{\rm star}(v_0)} \phi(x_0,x)=0.
\end{equation}
There is one equation for every vertex $v_0\in V(\cG)$; the
summation is extended over ${\rm star}(v_0)$, the set of vertices
of $\,\cG$ connected to $v_0$ by an edge (see Figure \ref{Fig:
star}); we write $x_0=x(v_0)$ and $x=x(v)$ and often suppress the
notational difference between the vertices $v$ of the graph and
the fields $x=x(v)$ assigned to them. Often, the function
$\phi=\phi(x_0,x;\alpha)$ is supposed to additionally depend on
some parameters $\alpha:E(\cG)\to\mathbb C$, assigned to the edges
of $\,\cG$.
\end{definition}
%------------------------------------------------------------------
\begin{figure}[htbp]
    \setlength{\unitlength}{16pt}
\begin{minipage}[t]{165pt}
\begin{picture}(9,10)(-1,-1)
\put(4,4){\circle*{0.3}} \put(4.2,8){\circle*{0.3}}
\put(7,5.9){\circle*{0.3}} \put(7.5,2.6){\circle*{0.3}}
\put(4,0.2){\circle*{0.3}} \put(0.4,3.9){\circle*{0.3}}
\path(4,4)(4.2,8) \path(4,4)(7,5.9) \path(4,4)(7.5,2.6)
\path(4,4)(4,0.2) \path(4,4)(0.4,3.9)
 \put(3.1,4.4){$x_0$}
 \put(7.4,5.9){$x_1$}
 \put(4.1,8.4){$x_2$}
 \put(0.2,4.3){$x_3$}
 \put(3.9,-0.4){$x_4$}
 \put(7.7,2.2){$x_5$}
\end{picture}
    \caption{Star of a vertex $x_0$ in the graph $\cG$.}
    \label{Fig: star}
\end{minipage}\hfill
\begin{minipage}[t]{170pt}
\begin{picture}(8,10)(-1,-1)
\put(4,4){\circle*{0.3}} \put(4.2,8){\circle*{0.3}}
\put(5,6){\circle{0.3}} \put(3,5.7){\circle{0.3}}
\put(7,5.9){\circle*{0.3}} \put(6,4.2){\circle{0.3}}
\put(7.5,2.6){\circle*{0.3}} \put(5.1,2.5){\circle{0.3}}
\put(4,0.2){\circle*{0.3}} \put(3,2.1){\circle{0.3}}
\put(0.4,3.9){\circle*{0.3}} \thicklines \path(4,4)(4.2,8)
\path(4,4)(7,5.9) \path(4,4)(7.5,2.6) \path(4,4)(4,0.2)
\path(4,4)(0.4,3.9)
 \dashline[+60]{0.3}(4.85,6)(3.15,5.75)
 \dashline[+60]{0.3}(3,5.55)(3.,2.25)
 \dashline[+60]{0.3}(3.15,2.1)(4.95,2.5)
 \dashline[+60]{0.3}(5.16,2.6)(5.95,4.05)
 \dashline[+60]{0.3}(5.95,4.35)(5.1,5.85)
% (6,4.2)=y0
% (5,6)=y1
% (3,5.7)=y2
% (3,2.1)=y3
% (5.1,2.5)=y4
 \put(3.15,3.5){$x_0$}
 \put(6.3,4.2){$y_1$}
 \put(7.3,5.9){$x_1$}
 \put(5.1,6.3){$y_2$}
 \put(4.1,8.3){$x_2$}
 \put(2.1,6){$y_3$}
 \put(-0.5,4.3){$x_3$}
 \put(2.4,1.6){$y_4$}
 \put(3.9,-0.3){$x_4$}
 \put(5.1,2.0){$y_5$}
 \put(7.5,2.1){$x_5$}
\end{picture}
   \caption{Face of $\cG^*$ dual to a vertex $x_0$ of $\cG$.}
   \label{Fig: dual face}
\end{minipage}
    \end{figure}
%------------------------------------------------------------------------

The notion of integrability of discrete Toda type systems is not
well established yet. We discuss here a definition based on the
notion of the discrete zero curvature representation which works
under an additional assumption about the graph $\cG$. Namely, it
has to come from a strongly regular polytopal cell decomposition
of an oriented surface.

We consider, in somewhat more detail, the dual graph (cell
decomposition) $\cG^*$. Each $\fe\in E(\cG)$ separates two faces
of $\cG$, which in turn correspond to two vertices of $\cG^*$. A
path between these two vertices is then declared the edge
$\fe^*\in E(\cG^*)$ dual to $\fe$. If one assigns a direction to
an edge $\fe\in E(\cG)$, then it will be assumed that the dual
edge $\fe^*\in E(\cG^*)$ is also directed, in a way consistent
with the orientation of the underlying surface, namely so that the
pair $(\fe,\fe^*)$ is positively oriented at its crossing point.
This orientation convention implies that $\fe^{**}=-\fe$. Finally,
the faces of $\cG^*$ are in a one-to-one correspondence with the
vertices of $\cG$: if $x_0\in V(\cG)$, and $x_1,\ldots,x_n\in
V(\cG)$ are its neighbors connected with $x_0$ by the edges
$\fe_1=(x_0,x_1),\ldots, \fe_n=(x_0,x_n)\in E(\cG)$, then the face
of $\cG^*$ dual to $x_0$ is bounded by the dual edges
$\fe_1^*=(y_1,y_2),\ldots, \fe_n^*=(y_n,y_1)$; see Figure
\ref{Fig: dual face}.

We will say that a discrete Toda type system on $\cG$ possesses a
discrete zero curvature representation if there is a collection of
matrices $L(\fe^*;\lambda)\in G[\lambda]$ from some loop group
$G[\lambda]$, associated to directed edges
$\fe^*\in\vec{E}(\cG^*)$ of the {\em dual graph} $\cG^*$, such
that:
\begin{itemize}
 \item the matrix $L(\fe^*;\lambda)=L(x_0,x,\alpha;\lambda)$ depends on the fields
$x_0$ and $x$ at the vertices of the edge $\fe=(x_0,x)\in E(\cG)$,
dual to the edge $\fe^*\in E(\cG^*)$, as well as on the parameter
$\alpha=\alpha(\fe)$;
 \item for any directed edge
$\fe^*=(y_1,y_2)$, if $-\fe=(y_2,y_1)$, then
\begin{equation}\label{zero curv cond inv}
L(-\fe,\lambda)=\big(L(\fe,\lambda)\big)^{-1};
\end{equation}
\item for any closed path of directed edges
\[
\fe^*_1=(y_1,y_2),\quad \fe^*_2=(y_2,y_3),\quad \ldots,\quad
\fe^*_n=(y_n,y_1),
\]
we have
\begin{equation}\label{zero curv cond}
L(\fe^*_n,\lambda)\cdots L(\fe^*_2,\lambda)L(\fe^*_1,\lambda)={\bf
1}.
\end{equation}
\end{itemize}
The matrix $L(\fe^*;\lambda)$ is interpreted as a transition
matrix {\em along} the edge $\fe^*\in E(\cG^*)$, that is, a
transition {\em across} the edge $\fe\in E(\cG)$.

Under conditions \eqref{zero curv cond inv}, (\ref{zero curv
cond}) one can define a {\em wave function} $\Psi: V(\cG^*)\to
G[\lambda]$ on the vertices of the dual graph $\cG^*$, by the
following requirement: for any directed edge $\fe^*=(y_1,y_2)$,
the values of the wave functions at its ends must be connected via
\begin{equation}\label{wave function evol}
\Psi(y_2,\lambda)=L(\fe^*,\lambda)\Psi(y_1,\lambda).
\end{equation}

For an arbitrary graph, the analytical consequences of the zero
curvature representation for a given collection of equations are
not clear. However, in the case of regular graphs, like the square
lattice or the regular triangular lattice, such a representation
may be used to determine conserved quantities for suitably defined
Cauchy problems, as well as to apply powerful analytical methods
for finding concrete solutions.

It was shown in \cite{A01} that discrete Toda type systems with
the following functions $\phi$ are integrable in the above sense:
\begin{eqnarray}
\phi(x_0,x;\alpha) & = & \frac{\alpha}{x-x_0}, \label{add rat}\\
\phi(x_0,x;\alpha) & = & \alpha\coth(x-x_0), \label{add hyp}\\
\phi(x_0,x;\alpha) & = & \log\frac{x-x_0+\alpha}{x-x_0-\alpha}, \label{mult rat}\\
\phi(x_0,x;\alpha) & = &
\log\frac{\sinh(x-x_0+\alpha)}{\sinh(x-x_0-\alpha)}. \label{mult
hyp}
\end{eqnarray}
See \cite{A01} for details about the admissible assignments of
edge parameters $\alpha$. Actually the discrete Toda system with
the functions (\ref{add hyp}) is reduced to the one with the
functions (\ref{add rat}) via the change of variables
$x\mapsto\exp(2x)$ and therefore actually does not need to be
considered separately.

\section{Quad-graphs and quad-equations}
\label{Sect quad-graphs}

\begin{figure}[htbp]
    \setlength{\unitlength}{16pt}
\begin{center}
\begin{picture}(8,9)(-1,0)
\put(4,4){\circle*{0.3}} \put(4.2,8){\circle*{0.3}}
\put(5,6){\circle{0.3}} \put(3,5.7){\circle{0.3}}
\put(7,5.9){\circle*{0.3}} \put(6,4.2){\circle{0.3}}
\put(7.5,2.6){\circle*{0.3}} \put(5.1,2.5){\circle{0.3}}
\put(4,0.2){\circle*{0.3}} \put(3,2.1){\circle{0.3}}
\put(0.4,3.9){\circle*{0.3}}
%\thicklines
\path(4.93,5.86)(4,4)\path(4.95,6.125)(4.2,8)\path(5.15,5.9875)(7,5.9)
% (5,6)=y1
\path(3.09,5.55)(4,4)\path(3.07,5.84)(4.2,8)\path(2.87,5.61)(0.4,3.9)
% (3,5.7)=y2
\path(3.07,2.24)(4,4)\path(2.87,2.19)(0.4,3.9)\path(3.07,1.96)(4,0.2)
% (3,2.1)=y3
\path(5.01,2.635)(4,4)\path(5.01,2.36)(4,0.2)\path(5.25,2.5)(7.5,2.6)
% (5.1,2.5)=y4
\path(5.85,4.185)(4,4)\path(6.09,4.1)(7.5,2.6)\path(6.09,4.35)(7,5.9)
% (6,4.2)=y0
 \put(3.1,4){$x_0$}
 \put(6.3,4.2){$y_1$}
 \put(7.3,5.9){$x_1$}
 \put(5.1,6.3){$y_2$}
 \put(4.1,8.3){$x_2$}
 \put(2.1,6){$y_3$}
 \put(-0.5,4.3){$x_3$}
 \put(2.4,1.6){$w_4$}
 \put(3.9,-0.3){$x_4$}
 \put(5.1,2.0){$y_5$}
 \put(7.5,2.1){$x_5$}
\end{picture}
   \caption{Faces of $\cD$ around the vertex $x_0$.}
   \label{Fig:flower}
\end{center}
    \end{figure}
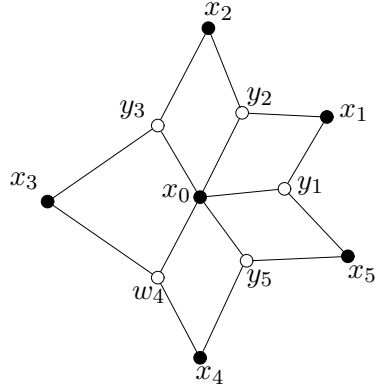

Although one can consider 2D integrable systems on very different
kinds of graphs on surfaces, there is one kind --- quad-graphs ---
supporting the most fundamental integrable systems.
\begin{definition}\label{Def: quad-graph}
A {\em quad-graph} $\cD$ is a strongly regular polytopal cell
decomposition of a surface with all quadrilateral faces.
\end{definition}
Quad-graphs are privileged because from an arbitrary strongly
regular polytopal cell decomposition $\cG$ one can produce a
certain quad-graph $\cD$, called the double of $\cG$. The {\it
double} $\cD$ is a quad-graph, constructed from $\cG$ and its dual
$\cG^*$ as follows.\index{double of a graph} The set of vertices
of the double $\cD$ is $V(\cD)=V(\cG)\sqcup V(\cG^*)$. Each pair
of dual edges, say $\fe=(x_0,x_1)\in E(\cG)$ and
$\fe^*=(y_1,y_2)\in E(\cG^*)$, defines a quadrilateral
$(x_0,y_1,x_1,y_2)$. These quadrilaterals constitute the faces of
a cell decomposition (quad-graph) $\cD$. Thus, a star of a vertex
$x_0\in V(\cG)$ generates a flower of adjacent quadrilaterals from
$F(\cD)$ around $x_0$; see Figure \ref{Fig:flower}. Let us stress
that edges of $\cD$ belong neither to $E(\cG)$ nor to $E(\cG^*)$.

%------------------------------------------------------------------
%------------------------------------------------------------------------

Quad-graphs $\cD$ coming as doubles are bipartite: the set
$V(\cD)$ may be decomposed into two complementary halves,
$V(\cD)=V(\cG)\sqcup V(\cG^*)$ (``black'' and ``white'' vertices),
such that the ends of each edge from $E(\cD)$ are of different
colors. Equivalently, any closed loop consisting of edges of $\cD$
has an even length.

The construction of the double can be reversed. Start with a
bipartite quad-graph $\cD$. For instance, any quad-graph embedded
in a plane or in an open disc is automatically bipartite. Any
bipartite quad-graph produces two dual polytopal (in general, no
more quadrilateral) cell decompositions $\cG$ and $\cG^*$, with
$V(\cG)$ containing all the ``black'' vertices of $\cD$ and
$V(\cG^*)$ containing all the ``white'' ones, and edges of $\cG$
(resp. of $\cG^*$) connecting ``black'' (resp. ``white'') vertices
along the diagonals of each face of $\cD$. The decomposition of
$V(\cD)$ into $V(\cG)$ and $V(\cG^*)$ is unique, up to
interchanging the roles of $\cG$ and $\cG^*$.

A privileged role played by the quad-graphs is reflected in the
privileged role played in the theory of discrete integrable
systems by the so called {\em quad-equations} supported by
quad-graphs.
\begin{definition} For a given bipartite quad-graph $\cD$, the
system of {\em quad-equations} for a function $x:V(\cD)\to\mathbb
C$ consists of equations of the type
\begin{equation}\label{eq:2d for 3leg}
Q(x_0,y_1,x_1,y_2)=0;
\end{equation}
see Figure \ref{Fig: quadrilateral bipartite}. There is one
equation for every face $(x_0,y_1,x_1,y_2)$ of $\cD$. The function
$Q$ is supposed to be {\em multi-affine}, i.e., a polynomial of
degree $\le 1$ in each argument, so that equation (\ref{eq:2d for
3leg}) is uniquely solvable for any of its arguments. Often, it is
supposed that the function $Q=Q(x_0,y_1,x_1,y_2;\alpha,\beta)$
additionally depends on some parameters usually assigned to the
edges of the quadrilaterals, $\alpha:E(\cD)\to\mathbb C$, so that
the opposite edges carry equal parameters:
$\alpha=\alpha(x_0,y_1)=\alpha(y_2,x_1)$ and
$\beta=\alpha(x_0,y_2)=\alpha(y_1,x_1)$.
\end{definition}

%------------------------------------------------------------------
\begin{figure}[htbp]
    \setlength{\unitlength}{0.05em}
\begin{minipage}[t]{200pt}
\begin{center}
\setlength{\unitlength}{0.05em}
\begin{picture}(200,220)(-100,-100)
 \put(-100,0){\circle*{10}}\put(100,0){\circle*{10}}
 \put(0,-80){\circle{10}} \put(0,80){\circle{10}}
 \thicklines
 \path(-100,0)(-3.9,-78.5)
 \path(-100,0)(-3.9,78.5)
 \path(100,0)(3.9,-78.5)
 \path(100,0)(3.9,78.5)
 \put(-127,  -5){$x_0$} \put(112, -5){$x_1$}
 \put(  -6,-105){$y_1$} \put( -6,97){$y_2$}
 % \put(-66,53){$\alpha_2$} \put(60,-60){$\alpha_2$}
 % \put(57,55){$\alpha_1$} \put(-65,-64){$\alpha_1$}
 \put(-10,0){$Q$}
\end{picture}
\caption{A quad-equation.}\label{Fig: quadrilateral bipartite}
\end{center}
\end{minipage}\hfill
\begin{minipage}[t]{160pt}
\begin{center}
\setlength{\unitlength}{0.05em}
\begin{picture}(200,220)(-100,-100)
 \put(-100,0){\circle*{10}}\put(100,0){\circle*{10}}
 \put(0,-80){\circle{10}} \put(0,80){\circle{10}}
 \path(-100,0)(-3.9,-78.5)
 \path(-100,0)(-3.9,78.5)
 {\thicklines\path(-100,0)(100,0)}
 \put(-127,  -5){$x_0$} \put(112, -5){$x_1$}
 \put(  -6,-105){$y_1$} \put( -6,97){$y_2$}
 \put(-66,55){$\psi$}\put(-65,-64){$\psi$}
 \put(0,15){$\phi$}
\end{picture}
\caption{Three-leg form of a quad-equation.}\label{Fig: 3leg}
\end{center}
\end{minipage}
    \end{figure}
%------------------------------------------------------------------------

There exists a fundamental and surprisingly simple notion of 3D
consistency of quad-equations which can be put into the basis of
the {\em integrability theory}, which has been done in \cite{BS02}
and \cite{N02}. The property of 3D consistency allows one, in
particular, to derive in an algorithmic way such basic
integrability attributes as discrete zero curvature
representations and B\"acklund transformations for quad-equations.
Moreover, this property has been put \cite{ABS03} into the basis
of a classification of integrable quad-equations which provided a
finite list of such equations known nowadays as the ``ABS list''.

\section{From quad-equations to discrete Toda type systems}
\label{Sect: Q to Toda}

The geometric relation of a given surface graph $\cG$ to its
double $\cD$, described in Section \ref{Sect quad-graphs}, leads
to a relation of discrete Toda type systems on $\cG$ to
quad-equations on $\cD$. The latter relation is based on a deep
and somewhat mysterious property of quad-equations which was
discovered in several examples in \cite{BS02}, was established for
all equations of the ABS list in \cite{ABS03}, and was proved for
all quad-equations with multi-affine functions $Q$ by V.~Adler,
see Exercise 6.16 in \cite{BS08}.

\begin{definition}\label{Def:3leg}
A quad-equation \eqref{eq:2d for 3leg} possesses a {\em three-leg
form} centered at the vertex $x_0$ if it is equivalent to the
equation
\begin{equation}\label{eq:3leg add}
\psi(x_0,y_1)-\psi(x_0,y_2)=\phi(x_0,x_1)
\end{equation}
with some functions $\psi, \phi$. The terms on the left-hand side
correspond to the ``short'' legs $(x_0,y_1), (x_0,y_2)\in E(\cD)$,
while the right-hand side corresponds to the ``long'' leg
$(x_0,x_1)\in E(\cG)$.
\end{definition}

Summation of quad-graph equations for the flower of quadrilaterals
adjacent to the ``black'' vertex $x_0\in V(\cG)$ (see Figure
\ref{Fig:flower}) immediately leads, due to the telescoping
effect, to the following statement.
\begin{theorem}\label{Th: Laplace for 3legs}

{\rm a)} Suppose that equation \eqref{eq:2d for 3leg} on a
bipartite quad-graph $\cD$ possesses a three-leg form. Then the
restriction of any solution $f:V(\cD)\to\mathbb C$ to the
``black'' vertices $V(\cG)$ satisfies the discrete Toda type
equations,
\begin{equation}\label{eq:Laplace for 3legs}
\sum_{x_k\in\,{\rm star}(x_0)} \phi(x_0,x_k)=0.
\end{equation}

{\rm b)} Conversely, given a solution $f:V(\cG)\to\mathbb C$ of
the Toda type equations \eqref{eq:Laplace for 3legs} on a simply
connected surface graph $\cG$, there exists a one-parameter family
of extensions $f:V(\cD)\to\mathbb C$ satisfying equation
\eqref{eq:2d for 3leg} on the double $\cD$. Such an extension is
uniquely determined by the value at one arbitrary vertex of
$V(\cG^*)$.
\end{theorem}

It was shown in \cite{BS02} that symmetric discrete Toda type
systems mentioned at the end of Section \ref{Sect: dToda} come,
through this construction, from the following integrable
quad-equations: the systems with legs (\ref{add rat}) and
(\ref{mult rat}) come from the $\delta=0$ and $\delta=1$ cases,
respectively, of the so called Q1 equation of the ABS list, which
reads
\[
\alpha(x_0y_1+x_1y_2)-\beta(x_0y_2+x_1y_1)-
  (\alpha-\beta)(x_0x_1+y_1y_2)+\delta\alpha\beta(\alpha-\beta)=0,
\]
while the system with legs (\ref{mult hyp}) comes from the
$\delta=0$ case of the so called Q3 equation of the ABS list,
which reads
\[
\sinh(\alpha)(x_0y_1+x_1y_2)-\sinh(\beta)(x_0y_2+x_1y_1)
      -\sinh(\alpha-\beta)(x_0x_1+y_1y_2)=0.
\]

\section{Triangular lattice}\label{s:7TL}

In Section \ref{Sect: Q to Toda} we established, for symmetric
discrete Toda systems on an arbitrary planar graph $\cG$, a
relation to integrable quad-equations on the double $\cD$. For
non-symmetric discrete relativistic Toda type systems, such a
relation remained unknown until recently, and it constitutes the
main new result of the present paper.

The non-symmetric discrete relativistic Toda type systems live on
the regular triangular lattice $\cT$ and cannot be directly
generalized to arbitrary graphs. Therefore, we introduce now the
specific notation tailored for the regular triangular lattice. The
double of $\cT$ is the quad-graph $\cK$ known as the {\em dual
kagome lattice} (drawn on Figure \ref{fig:L3} in dashed lines).
The latter graph has vertices of two kinds, black vertices of
valence 6 and white vertices of valence 3, and edges of three
types, all edges of each type being parallel. The quadrilateral
faces of the dual kagome lattice are of three different types. We
will denote them by type I, II, and III, according to Figure
\ref{Fig: quads}.

%-----------------------------------------------------------------
\setcircle{5}
\def\tmpga(#1,#2){\put(#1,#2){\path(0,0)(0,100)(100,0)(0,0)
 \Black(0,0) \Black(100,0) \Black(0,100) \White(34,34)
 \dashline{4}(0,0)(30,30) \dashline{4}(0,100)(32,39)
 \dashline{4}(100,0)(39,32)}}
\def\tmpgb(#1,#2){\put(#1,#2){\White(-34,-34)
 \dashline{4}(0,0)(-30,-30) \dashline{4}(0,-100)(-32,-39)
 \dashline{4}(-100,0)(-39,-32)}}
 %-----------------------------------------------------------------
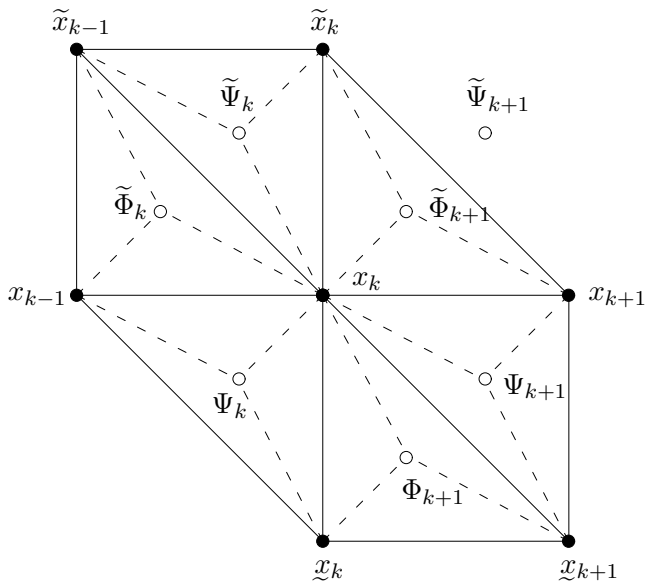
\begin{figure}[htbp]
\begin{center}
\setlength{\unitlength}{0.085em}
\begin{picture}(300,240)(-150,-120)
 \tmpga(-100,0)\tmpga(0,0)\tmpga(0,-100)
 \tmpgb( 100,0)\tmpgb(0,0)\tmpgb(0, 100)
  \path(-100,0)(0,-100) \path(100,0)(100,-100) \path(0,100)(-100,100)
  \put(-110,108){$\wx_{k-1}$} \put(-5,108){$\wx_k$}
  \put(-128,-2){$x_{k-1}$} \put(12,4){$x_k$} \put(108,-2){$x_{k+1}$}
  \put(-5,-113){$\undertilde{x}_k$} \put(95,-113){$\undertilde{x}_{k+1}$}
  \put(-45,-48){$\Psi_k$}
  \put( 32,-83){$\Phi_{k+1}$}
  \put( 73,-40){$\Psi_{k+1}$}
  \put( 43, 32){$\widetilde\Phi_{k+1}$}
  \put(-42, 77){$\widetilde\Psi_k$}
  \put(-85, 33){$\widetilde\Phi_k$}
  \put( 58, 77){$\widetilde\Psi_{k+1}$}\White(66,66)
\end{picture}
\caption{Fields and wave functions on the triangular lattice}
\label{fig:L3}
\end{center}
\end{figure}
%-----------------------------------------------------------------

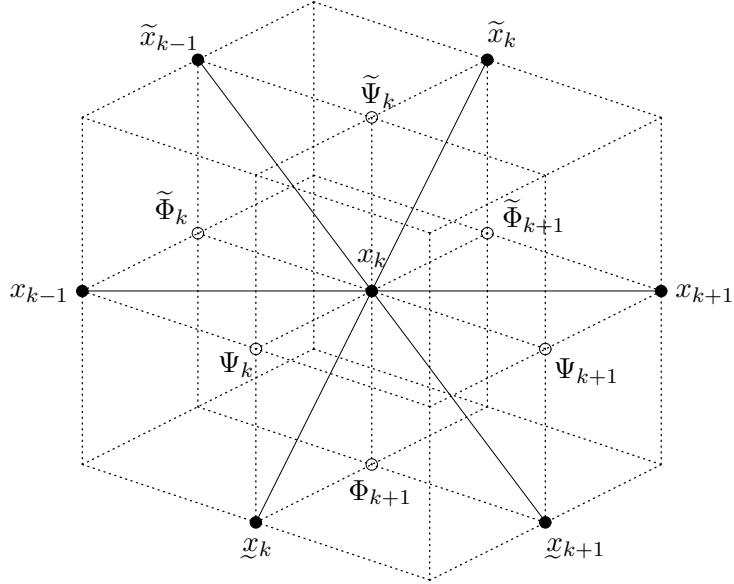
\begin{figure}[htb]
\begin{center}
\setlength{\unitlength}{0.1em} \setcircle{4}
%-----------------------------------
\begin{picture}(240,200)(-20,-40)
 \multiput(0,0)(0,60){3}{
  \multiput(0,0)(40,20){3}{\dottedline{2}(0,0)(120,-40)}
  \multiput(0,0)(60,-20){3}{\dottedline{2}(0,0)(80,40)}}
 \multiput(0,0)(40,20){3}{
  \multiput(0,0)(60,-20){3}{\dottedline{2}(0,0)(0,120)}}

 \Black(  0, 60)\put(-25,58){$x_{k-1}$}
 \Black(100, 60)\put(95,70){$x_k$}
 \Black(200, 60)\put(205,58){$x_{k+1}$}
 \Black( 60,-20)\put( 54,-30){$\undertilde{x}_k$}
 \Black(140,140)\put(140,146){$\wx_k$}
 \Black( 40,140)\put( 20,145){$\wx_{k-1}$}
 \Black(160,-20)\put(159,-30){$\undertilde{x}_{k+1}$}

 \White( 60, 40)\put( 47,32){$\Psi_k$}
 \White(160, 40)\put(163,30){$\Psi_{k+1}$}
 \White(100,120)\put(96,125){$\widetilde\Psi_k$}

 \White(40,80)\put(25,85){$\widetilde\Phi_k$}
 \White(140,80)\put(145,82){$\widetilde\Phi_{k+1}$}
 \White(100,0)\put(92,-12){$\Phi_{k+1}$}

% \put(64,68){$\a_{2,-1}$} \put(103,102){$\a_3$}
% \put(66,50){$\a_1$} \put(125,43){$\a_2$}
 \path(0,60)(200,60)\path(40,140)(160,-20)\path(60,-20)(140,140)
\end{picture}
%-----------------------------------
\caption{Embedding of the triangular lattice and the dual kagome
lattice into $\mathbb Z^3$} \label{fig:star6}
\end{center}
\end{figure}

The dual kagome lattice can be realized as a quad-surface in
$\mathbb Z^3$, so that the three types of quadrilaterals are
realized as elementary squares of $\mathbb Z^3$ parallel to the
three coordinate planes (this is easy to see directly but follows
also from the general theory of quasi-crystallic quad-graphs in
\cite{BMS05}). In this realization, the black vertices of $\cK$,
that is, the vertices of $\cT$, are the points
$(i_1,i_2,i_3)\in\mathbb Z^3$ lying in the plane $i_1+i_2+i_3=0$,
while the white vertices of $\cK$ are the points of $\mathbb Z^3$
lying in the planes $i_1+i_2+i_3=1$ (the vertices $\Psi$) and
$i_1+i_2+i_3=-1$ (the vertices $\Phi$). See Figure
\ref{fig:star6}.

%-----------------------------------------------------------------
%-----------------------------------------------------------------

\section{Discrete relativistic Toda type system from
quad-equations on $\cK$} \label{Sect main}

We now formulate the main result of the present paper.

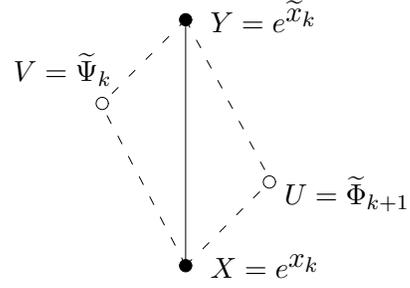
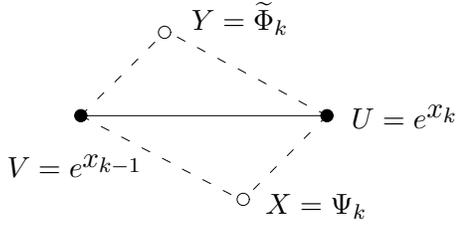
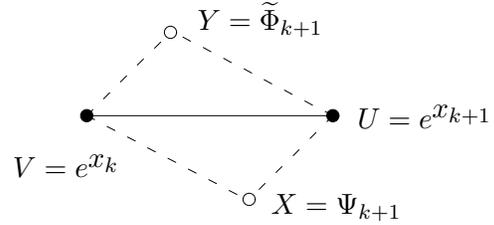
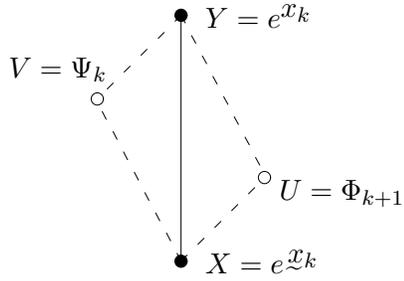
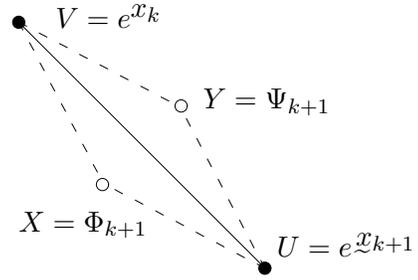
\begin{figure}[htbp]
    \setlength{\unitlength}{0.085em}
\centering \subfigure[North-western quadrilateral of type III] {
\begin{picture}(160,150)(-25,-120)
 \Black(0,0)\Black(100,-100)\White(34,-66)\White(66,-34)
 \path(0,0)(100,-100)
 \dashline{4}(0,0)(61,-32)
 \dashline{4}(0,0)(32,-61)
 \dashline{4}(100,-100)(39,-68)
 \dashline{4}(100,-100)(68,-39)
 \put(15,-2){$V=\eto{\wx_{k-1}}$}
 \put(75,-35){$Y=\widetilde{\Psi}_k$}
 \put(105,-95){$U=\eto{x_k}$}
 \put(0,-85){$X=\widetilde{\Phi}_k$}
\end{picture}
\label{Fig: north-west quad}}\qquad\qquad
 \subfigure[Northern quadrilateral of type I] {
\begin{picture}(160,150)(-75,-20)
 \Black(0,100)\Black(0,0)\White(-34,66)\White(34,34)
 \path(0,0)(0,100)
 \dashline{4}(0,0)(30,30)
 \dashline{4}(0,0)(-32,61)
 \dashline{4}(0,100)(-30,70)
 \dashline{4}(0,100)(32,39)
 \put(10,-5){$X=\eto{x_k}$}
 \put(40,25){$U=\widetilde\Phi_{k+1}$}
 \put(10,95){$Y=\eto{\wx_k}$}
 \put(-70,75){$V=\widetilde\Psi_k$}
\end{picture}
\label{Fig: north quad}}\\
\subfigure[Western quadrilateral of type II] {
\begin{picture}(160,100)(-25,-50)
 \Black(0,0)\Black(100,0)\White(34,34)\White(66,-34)
 \path(0,0)(100,0)
 \dashline{4}(0,0)(30,30)
 \dashline{4}(0,0)(61,-32)
 \dashline{4}(100,0)(39,32)
 \dashline{4}(100,0)(70,-30)
 \put(75,-40){$X=\Psi_k$}
 \put(45,35){$Y=\widetilde\Phi_k$}
 \put(110,-5){$U=\eto{x_k}$}
 \put(-30,-25){$V=\eto{x_{k-1}}$}
\end{picture}
\label{Fig: west quad}}\qquad\qquad
 \subfigure[Eastern quadrilateral of type II] {
\begin{picture}(160,100)(-25,-50)
 \Black(0,0)\Black(100,0)\White(34,34)\White(66,-34)
 \path(0,0)(100,0)
 \dashline{4}(0,0)(30,30)
 \dashline{4}(0,0)(61,-32)
 \dashline{4}(100,0)(39,32)
 \dashline{4}(100,0)(70,-30)
 \put(75,-40){$X=\Psi_{k+1}$}
 \put(45,35){$Y=\widetilde\Phi_{k+1}$}
 \put(110,-5){$U=\eto{x_{k+1}}$}
 \put(-30,-25){$V=\eto{x_{k}}$}
\end{picture}
\label{Fig: east quad}}\\
\subfigure[Southern quadrilateral of type I] {
\begin{picture}(160,150)(-75,-20)
 \Black(0,100)\Black(0,0)\White(-34,66)\White(34,34)
 \path(0,0)(0,100)
 \dashline{4}(0,0)(30,30)
 \dashline{4}(0,0)(-32,61)
 \dashline{4}(0,100)(-30,70)
 \dashline{4}(0,100)(32,39)
 \put(10,-5){$X=\eto{\undertilde{x}_k}$}
 \put(40,25){$U=\Phi_{k+1}$}
 \put(10,95){$Y=\eto{x_k}$}
 \put(-70,75){$V=\Psi_k$}
\end{picture}
\label{Fig: south quad}}\qquad\qquad
 \subfigure[South-eastern quadrilateral of type III] {
\begin{picture}(160,150)(-25,-120)
 \Black(0,0)\Black(100,-100)\White(34,-66)\White(66,-34)
 \path(0,0)(100,-100)
 \dashline{4}(0,0)(61,-32)
 \dashline{4}(0,0)(32,-61)
 \dashline{4}(100,-100)(39,-68)
 \dashline{4}(100,-100)(68,-39)
 \put(15,-2){$V=\eto{x_k}$}
 \put(75,-35){$Y={\Psi}_{k+1}$}
 \put(105,-95){$U=\eto{\undertilde{x}_{k+1}}$}
 \put(0,-85){$X={\Phi}_{k+1}$}
\end{picture}
\label{Fig: south-east quad}}
 \caption{Notation for single quadrilaterals of the dual kagome
 lattice around the vertex $x_k$}
 \label{Fig: quads}
    \end{figure}

\begin{theorem}\label{Th main}
Each discrete relativistic Toda type system is a restriction to
the triangular lattice $\cT$ of a certain 3D consistent system of
quad-equations on the dual kagome lattice $\cK$ considered as a
quad-surface in $\mathbb Z^3$.
\end{theorem}
{\bf Proof} of this theorem is obtained by a direct case-by-case
construction of the corresponding systems of quad-equations (see,
however, about the unifying ``master system'' in Section \ref{Sect
master}). For the lack of space, these systems are given below not
for all discrete relativistic Toda systems, but for four of them
only, namely, for (\ref{dRTL+ lq New introd}), (\ref{dRTL+ l New
introd}), (\ref{dRTL+ m New introd}), and (\ref{dRTL+ dual New
introd}). Details for other systems can be found in \cite{B08}.
The systems are specified by giving the quad-equations explicitly
for each type of quadrilateral faces separately in notation of
Figure \ref{Fig: quads}. One has to: a) find the three-leg forms,
centered at $x_k$, of quad-equations for all six quadrilaterals
around $x_k$ and then check that adding these three-leg forms
results in the corresponding discrete Toda equation, and b) check
the 3D consistency of the quad-equations. All this is a matter of
direct computations which are easy enough to perform by hands but
are better delegated to a symbolic manipulator like Maple or
Mathematica. \hfill $\Box$

%------------------------------------------------------------------
%------------------------------------------------------------------------
\medskip

\noindent {\bf System (\ref{dRTL+ lq New introd}).}\quad 3D
consistent system of quad-equations:
\begin{align}
& \alpha h\lambda(XY+UV)-\alpha h(\alpha-h)\lambda^2XU-
\left((\alpha-h)\lambda^2+h\right)XV+\alpha\lambda^2YV=0,  \tag{I}\\
& \alpha\lambda(XY+UV)-XV+\lambda^2YV=0,  \tag{II} \\
& \alpha(\alpha-h)\lambda(XY+UV)+\alpha h(\alpha-h)XU+
\left((\alpha-h)\lambda^2+h\right)XV-\alpha YV=0.\quad\tag{III}
\end{align}
The three-leg forms of these equations (centered at $x_k$) read:
\begin{align}
& \left(\alpha \eto{\wx_{k}\nm
x_k}-\left(\alpha-h\right)\right)\cdot
\frac{h\eto{x_{k}}+\lambda\widetilde{\Psi}_{k}}
{\widetilde{\Psi}_{k}-\left(\alpha-h\right)\lambda\eto{x_{k}}}\cdot
\frac{\eto{x_{k}}}{\eto{x_{k}}-\alpha\lambda\widetilde{\Phi}_{k+1}}
=\frac{h}{\lambda},  \tag{N}\\
& \frac{1}{1+\alpha^2 \eto{x_{k+1}\nm x_{k}}}\cdot
\frac{\eto{x_{k}}-\alpha\lambda\widetilde{\Phi}_{k+1}}{\eto{x_k}}\cdot
\frac{\lambda \eto{x_{k}}+\alpha\Psi_{k+1}}{\eto{x_{k}}}
=\lambda,  \tag{E}\\
& \left(1+\alpha\left(\alpha-h\right)\eto{\undertilde{x}_{k+1}\nm
x_{k}}\right)\cdot
\frac{\eto{x_{k}}}{\lambda\eto{x_{k}}+\alpha\Psi_{k+1}}\cdot
\frac{\lambda \eto{x_{k}}+h\Phi_{k+1}}
{\eto{x_{k}}-\left(\alpha-h\right)\lambda\Phi_{k+1}}
=1,  \tag{SE}\\
& \frac{1}{\alpha \eto{x_{k}\nm
\undertilde{x}_{k}}-\left(\alpha-h\right)}\cdot
\frac{\eto{x_{k}}-\left(\alpha-h\right)\lambda\Phi_{k+1}}{\lambda
\eto{x_{k}}+h\Phi_{k+1}} \cdot
\frac{\Psi_{k}-\alpha\lambda\eto{x_{k}}}{\Psi_{k}}
=\frac{\lambda}{h},  \tag{S}\\
& \left(1+\alpha^2 \eto{x_{k}\nm x_{k-1}}\right)\cdot
\frac{\Psi_{k}}{\Psi_{k}-\alpha\lambda \eto{x_{k}}}\cdot
\frac{\widetilde{\Phi}_{k}}{\alpha\eto{x_{k}}+\lambda\widetilde{\Phi}_{k}}
=\frac{1}{\lambda},  \tag{W}\\
& \frac{1} {1+\alpha\left(\alpha-h\right)
\eto{x_{k}\nm\wx_{k-1}}}\cdot
\frac{\alpha\eto{x_{k}}+\lambda\widetilde{\Phi}_{k}}{\widetilde{\Phi}_{k}}\cdot
\frac{\widetilde{\Psi}_{k}-\left(\alpha-h\right)\lambda\eto{x_{k}}}
{h\eto{x_{k}}+\lambda\widetilde{\Psi}_{k}}=1.\; \tag{NW}
\end{align}
Multiplying these equations leads to (\ref{dRTL+ lq New introd}).
\medskip

\noindent {\bf System (\ref{dRTL+ l New introd}).}\quad 3D
consistent system of quad-equations:
\begin{align}
& h(XY+UV)+YV-\left(1-h\lambda\right)XV+h^2XU=0, \tag{I}\\
& \alpha (XY+UV)+YV-\left(1-\alpha\lambda\right)XV+\alpha^2XU=0, \tag{II}\\
& (h-\alpha)(XY+UV)+(1-\alpha\lambda)YV-
(1-h\lambda)XV+  \nonumber\\
& \qquad\qquad +h^2(1-\alpha\lambda)XU-\alpha^2(1-h\lambda)YU=0.
\tag{III}
\end{align}
Three-leg forms of these equations, centered at $x_k$:
\begin{align}
& \eto{\wx_{k}\nm x_{k}}+
\frac{h\widetilde{\Phi}_{k+1}}{\eto{x_{k}}}
-\frac{(1-h\lambda)\widetilde{\Psi}_k}{\widetilde{\Psi}_{k}+h\eto{x_{k}}}
=0,  \tag{N}\\
& -\alpha\eto{x_{k+1}\nm
x_{k}}-\frac{\widetilde{\Phi}_{k+1}}{\eto{x_{k}}}+
\frac{(1-\alpha\lambda)\Psi_{k+1}}{\eto{x_{k}}+\alpha\Psi_{k+1}}
=0,  \tag{E}\\
& \frac{(\alpha-h)\eto{\undertilde{x}_{k+1}\nm x_k}}
{1-h\alpha\eto{\undertilde{x}_{k+1}\nm x_{k}}}-
\frac{(1-\alpha\lambda)\Psi_{k+1}}{\eto{x_{k}}+\alpha\Psi_{k+1}}+
\frac{(1-h\lambda)\Phi_{k+1}}{\eto{x_{k}}+h\Phi_{k+1}}
=0,  \tag{SE}\\
& -\eto{x_{k}\nm\undertilde{x}_{k}}+
\frac{(1-h\lambda)\eto{x_{k}}}{\eto{x_{k}}+h\Phi_{k+1}}
-\frac{h\eto{x_{k}}}{\Psi_{n}}
=0,  \tag{S}\\
& \alpha \eto{x_{k}-x_{k-1}}+\frac{\eto{x_{k}}}{\Psi_{k}}-
\frac{(1-\alpha\lambda)\eto{x_{k}}}
{\widetilde{\Phi}_{k}+\alpha\eto{x_{k}}}
=0,  \tag{W}\\
& -\frac{(\alpha-h)\eto{x_{k}\nm
\wx_{k-1}}}{1-h\alpha\eto{x_{k}\nm
\wx_{k-1}}}+\frac{(1-\alpha\lambda)\eto{x_{k}}}
{\widetilde{\Phi}_{k}+\alpha\eto{x_{k}}}-
\frac{(1-h\lambda)\eto{x_{k}}}{\widetilde{\Psi}_{k}+h\eto{x_{k}}}
=0.   \tag{NW}
\end{align}
Adding these equations leads to (\ref{dRTL+ l New introd}).
\medskip

\noindent {\bf System (\ref{dRTL+ m New introd}).}\quad 3D
consistent system of quad-equations:
\begin{align}
& h\lambda (XY+UV)-h\lambda^2XU-\left(\lambda^2+h\right)XV+
\lambda^2YV =0, \tag{I}\\
& \alpha\lambda (XY+UV)
-\alpha\lambda^2XU-\left(\lambda^2+\alpha\right)XV
+\lambda^2YV =0,  \tag{II} \\
& (h-\alpha)\lambda (XY+UV)-h(\lambda^2+\alpha)XU-
(\lambda^2+h)XV+\nonumber\\
& \qquad+\alpha(\lambda^2+h)YU+(\lambda^2+\alpha)YV  =0. \tag{III}
\end{align}
Three-leg forms of these equations, centered at $x_k$:
\begin{align}
& \left(\eto{\wx_{k}\nm x_{k}}-1\right)\cdot
\frac{\lambda\widetilde{\Psi}_{k}+h\eto{x_{k}}}
{\widetilde{\Psi}_{k}-\lambda\eto{x_{k}}}\cdot
\frac{\eto{x_{k}}}{\eto{x_{k}}-\lambda\widetilde{\Phi}_{k+1}}=\frac{h}{\lambda}\,,
\tag{N}\\
& \frac{1}{1+\alpha\eto{x_{k+1}\nm x_k}}\cdot
\frac{\eto{x_{k}}-\lambda\widetilde{\Phi}_{k+1}}{\eto{x_{k}}}\cdot
\frac{\lambda\eto{x_{k}}+\alpha\Psi_{k+1}}{\eto{x_{k}}-\lambda\Psi_{k+1}}
=\lambda,  \tag{E}\\
& \frac{1+\alpha\eto{\undertilde{x}_{k+1}\nm
x_k}}{1+h\eto{\undertilde{x}_{k+1}-x_{k}}}\cdot
\frac{\eto{x_{k}}-\lambda\Psi_{k+1}}{\lambda
\eto{x_{k}}+\alpha\Psi_{k+1}}\cdot
\frac{\lambda\eto{x_{k}}+h\Phi_{k+1}}{\eto{x_{k}}-\lambda\Phi_{k+1}}
=1,  \tag{SE}\\
& \frac{1}{\eto{x_{k}\nm\undertilde{x}_{k}}-1}\cdot
\frac{\eto{x_{k}}-\lambda\Phi_{k+1}}{\lambda
\eto{x_{k}}+h\Phi_{k+1}}\cdot \frac{\Psi_{k}-\lambda
\eto{x_{k}}}{\Psi_{k}}
=\frac{\lambda}{h}\,,  \tag{S}\\
& \left(1+\alpha \eto{x_{n}\nm x_{n-1}}\right)\cdot
\frac{\Psi_{k}}{\Psi_{k}-\lambda \eto{x_{k}}}\cdot
\frac{\widetilde{\Phi}_{k}-\lambda\eto{x_{k}}}
{\lambda\widetilde{\Phi}_{k}+\alpha \eto{x_{k}}}
=\frac{1}{\lambda}\,,  \tag{W}\\
& \frac{1+h\eto{x_{k}\nm\wx_{k-1}}}{1+\alpha
\eto{x_{k}-\wx_{k-1}}}\cdot
\frac{\lambda\widetilde{\Phi}_{k}+\alpha\eto{x_{k}}}
{\widetilde{\Phi}_{k}-\lambda\eto{x_{k}}}\cdot
\frac{\widetilde{\Psi}_{k}-\lambda
\eto{x_{k}}}{\lambda\widetilde{\Psi}_{k}+h\eto{x_{k}}}=1. \tag{NW}
\end{align}
Multiplying these equations leads to (\ref{dRTL+ m New introd}).
\medskip

\noindent {\bf System (\ref{dRTL+ dual New introd}).}\quad 3D
consistent system of quad-equations:
\begin{align}
& h(XY-XU-YV+UV)-(1-h\lambda)(X-Y)- h\lambda(U-V)-h\lambda^2=0,
\tag{I}\\
& \alpha(XY-XU-YV+UV)-(1-\alpha\lambda)(X-Y)-
\alpha\lambda(U-V)-\alpha\lambda^2=0, \tag{II} \\
& (h-\alpha)(XY+UV)-h(1-2\alpha\lambda)
(XU+YV)+\alpha(1-2h\lambda)(XV+YU)-
\nonumber\\
& \qquad-\left(1-(\alpha+h)\lambda\right)(X-Y)-
(h-\alpha)\lambda(U-V)-(h-\alpha)\lambda^2=0. \tag{III}
\end{align}
Three-leg forms of these equations, centered at $x_k$:
\begin{align}
& (\wx_{k}-x_{k})\cdot
\frac{1+h(x_{k}-\widetilde{\Psi}_{k}-\lambda)}
{x_{k}-\widetilde{\Psi}_{k}+\lambda}\cdot
\frac{1}{x_{k}-\widetilde{\Phi}_{k+1}-\lambda}
=-h,  \tag{N}\\
& \frac{1}{1+\alpha(x_{k+1}-x_{k})}\cdot
(x_{k}-\widetilde{\Phi}_{k+1}-\lambda)\cdot
\frac{1-\alpha(x_{k}-\Psi_{k+1}+\lambda)}
{x_{k}-\Psi_{k+1}-\lambda}
=1,  \tag{E}\\
& \frac{1+\alpha(\undertilde{x}_{k+1}-x_{k})}
{1+h(\undertilde{x}_{k+1}-x_{k})}\cdot
\frac{x_{k}-\Psi_{k+1}-\lambda}
{1-\alpha(x_{k}-\Psi_{k+1}+\lambda)}\cdot
\frac{1-h(x_{k}-\Phi_{k+1}+\lambda)} {x_{k}-\Phi_{k+1}-\lambda}
=1,  \tag{SE}\\
& \frac{1}{x_{k}-\undertilde{x}_{k}}\cdot
\frac{x_{k}-\Phi_{k+1}-\lambda}{1-h(x_{k}-\Phi_{k+1}+\lambda)}\cdot
(x_{k}-\Psi_{k}+\lambda)
=-\frac{1}{h}\,,  \tag{S}\\
& (1+\alpha\left(x_{k}-x_{k-1}\right))\cdot
\frac{1}{x_{k}-\Psi_{k}+\lambda}\cdot
\frac{x_{k}-\widetilde{\Phi}_{k}+\lambda}
{1+\alpha(x_{k}-\widetilde{\Phi}_{k}-\lambda)}
=1,  \tag{W}\\
& \frac{1+h(x_{k}-\wx_{k-1})}{1+\alpha(x_{k}-\wx_{k-1})}\cdot
\frac{1+\alpha(x_{k}-\widetilde{\Phi}_{k}-\lambda)}
{x_{k}-\widetilde{\Phi}_{k}+\lambda}\cdot
\frac{x_{k}-\widetilde{\Psi}_{k}+\lambda}
{1+h(x_{k}-\widetilde{\Psi}_{k}-\lambda)}=1.  \tag{NW}
\end{align}
Multiplying these equations leads to (\ref{dRTL+ dual New
introd}).

\section{The master system}\label{Sect master}
It turns out that all the 3D consistent systems of quad-equations
leading to non-symmetric discrete relativistic Toda systems (those
given above and those omitted for the space reasons), as well as
the systems Q1 and Q3$_{\delta=0}$ leading to the symmetric
discrete relativistic Toda systems are particular or limiting
cases of one multi-parametric system. Thus this latter system can
be seen as the master one behind the whole theory of the
relativistic Toda systems of the type (\ref{dRTL gen intro}) (with
discrete time) and (\ref{RTL gen intro}) (with continuous time).
\medskip

{\bf Master system} of quad-equations:
\begin{align}
& (\delta-\beta\gamma)\lambda(
XY+UV)+\beta\delta(\lambda^2-\gamma)XU+(\beta\lambda^2-\delta)XV+
 \nonumber\\
& \qquad +\gamma(\beta\lambda^2-\delta)YU+(\lambda^2-\gamma)YV=0,
 \tag{I}\\
& (\eta-\gamma\epsilon)\lambda
(XY+UV)+(\lambda^2-\gamma)XU+(\epsilon\lambda^2-\eta)XV+
\nonumber\\
& \qquad+\gamma(\epsilon\lambda^2-\eta)YU+
\epsilon\eta(\lambda^2-\gamma)YV=0,
\tag{II}\\
& (\beta\eta-\delta\epsilon)\lambda
(XY+UV)-\beta\delta(\epsilon\lambda^2-\eta)XU+
\epsilon\eta(\beta\lambda^2-\delta)XV+
\nonumber\\
&\qquad+(\beta\lambda^2-\delta)YU-(\epsilon\lambda^2-\eta)YV=0.
\tag{III}
\end{align}
Three-leg forms of these equations, centered at $x_k$:
\begin{align}
& \frac{\eto{\wx_{k}}+\beta\eto{x_{k}}}
  {\gamma\eto{\wx_{k}}+\delta\eto{x_{k}}}\cdot
 \frac{\delta\eto{x_{k}}+\lambda\widetilde{\Psi}_{k}}
  {\beta\lambda\eto{x_{k}}+\widetilde{\Psi}_{k}}\cdot
 \frac{\lambda\eto{x_{k}}-\gamma\widetilde{\Phi}_{k+1}}
  {\eto{x_{k}}-\lambda\widetilde{\Phi}_{k+1}}=1,\tag{N}\\
& \frac{\gamma\eto{x_{k+1}}+\eta\eto{x_{k}}}
  {\eto{x_{k+1}}+\epsilon\eto{x_{k}}}\cdot
 \frac{\eto{x_{k}}-\lambda\widetilde{\Phi}_{k+1}}
  {\lambda\eto{x_{k}}-\gamma\widetilde{\Phi}_{k+1}}\cdot
 \frac{\epsilon\lambda\eto{x_{k}}+\Psi_{k+1}}
  {\eta\eto{x_{k}}+\lambda\Psi_{k+1}}=1, \tag{E}\\
& \frac{\beta\eto{\undertilde{x}_{k+1}}-\epsilon\eto{x_{k}}}
   {\delta\eto{\undertilde{x}_{k+1}}-\eta\eto{x_{k}}}\cdot
 \frac{\eta\eto{x_{k}}+\lambda\Psi_{k+1}}
  {\epsilon\lambda\eto{x_{k}}+\Psi_{k+1}}\cdot
 \frac{\lambda\eto{x_{k}}+\delta\Phi_{k+1}}
  {\eto{x_{k}}+\beta\lambda\Phi_{k+1}}=1, \tag{SE}\\
& \frac{\gamma\eto{x_{k}}+\delta\eto{\undertilde{x}_{k}}}
   {\eto{x_{k}}+\beta\eto{\undertilde{x}_{k}}}\cdot
 \frac{\eto{x_{k}}+\beta\lambda\Phi_{k+1}}
  {\lambda\eto{x_{k}}+\delta\Phi_{k+1}}\cdot
 \frac{\lambda\eto{x_{k}}-\Psi_{k}}
  {\gamma\eto{x_{k}}-\lambda\Psi_{k}}=1, \tag{S}\\
& \frac{\eto{x_{k}}+\epsilon \eto{x_{k-1}}}
   {\gamma\eto{x_{k}}+\eta\eto{x_{k-1}}}\cdot
  \frac{\gamma\eto{x_{k}}-\lambda\Psi_{k}}
   {\lambda\eto{x_{k}}-\Psi_{k}}\cdot
  \frac{\lambda\eto{x_{k}}+\eta\widetilde{\Phi}_{k}}
   {\eto{x_{k}}+\epsilon\lambda\widetilde{\Phi}_{k}}=1, \tag{W}\\
& \frac{\delta\eto{x_{k}}-\eta\eto{\wx_{k-1}}}
   {\beta\eto{x_{k}}-\epsilon\eto{\wx_{k-1}}}\cdot
  \frac{\eto{x_{k}}+\epsilon\lambda\widetilde{\Phi}_{k}}
   {\lambda\eto{x_{k}}+\eta\widetilde{\Phi}_{k}}\cdot
  \frac{\beta\lambda\eto{x_{k}}+\widetilde{\Psi}_{k}}
   {\delta\eto{x_{k}}+\lambda\widetilde{\Psi}_{k}}=1.  \tag{NW}
\end{align}
Multiplying these three-leg forms leads to the following general
equation of the discrete relativistic Toda type:
\begin{multline} \label{generalToda}
\frac{\eto{\wx_{k}\nm x_{k}}+\beta}
 {\gamma\eto{\wx_{k}\nm x_{k}}+\delta}\cdot
\frac{\gamma\eto{x_{k}\nm\undertilde{x}_{k}}+\delta}
 {\eto{x_{k}\nm\undertilde{x}_{k}}+\beta} \\
\cdot \frac{\gamma\eto{x_{k+1}\nm x_{k}}+\eta}
 {\eto{x_{k+1}\nm x_{k}}+\epsilon}\cdot
\frac{\eto{x_{k}\nm x_{k-1}}+\epsilon}
 {\gamma\eto{x_{k}\nm x_{k-1}}+\eta}\qquad\\
\cdot \frac{\beta\eto{\undertilde{x}_{k+1}\nm x_{k}}-\epsilon}
 {\delta\eto{\undertilde{x}_{k+1}\nm x_{k}}-\eta}\cdot
\frac{\delta\eto{x_{k}\nm \wx_{k-1}}-\eta}
 {\raisebox{-4pt}{$\beta\eto{x_{k}\nm\wx_{k-1}}-\epsilon$}}=1.\quad
\end{multline}
This equation has five parameters
$\beta,\gamma,\delta,\epsilon,\eta$. Actually there are only four,
because of homogeneity: if $\gamma\neq 0$, we can set $\gamma=1$
by replacing $\delta,\eta$ through $\delta/\gamma$, $\eta/\gamma$,
respectively. Moreover, we could eliminate two further parameters
by shifts of the form $x_k(t)\to x_k(t)+Ak+Bt$, which keep the
form of the equation invariant.

It is not difficult to find out the special values of parameters
which lead to all the discrete relativistic Toda type equations
listed in Section \ref{Sect dRTL}. In particular, the value
$\gamma=1$ leads to the most general symmetric equation (\ref{dRTL
cq New introd}), with further degenerations to (\ref{dRTL c2 New
introd}), (\ref{dRTL c New introd}), (\ref{dRTL c3 New introd}).
The value $\gamma=0$ is of the primary interest for the aims of
the present paper, since it leads to
\begin{equation} \label{generalToda nonsymm}
\frac{\eto{\wx_{k}\nm x_{k}}+\beta}
 {\eto{x_{k}\nm\undertilde{x}_{k}}+\beta}\cdot
\frac{\eto{x_{k}\nm x_{k-1}}+\epsilon}
 {\eto{x_{k+1}\nm x_{k}}+\epsilon}\cdot
\frac{\beta\eto{\undertilde{x}_{k+1}\nm x_{k}}-\epsilon}
 {\delta\eto{\undertilde{x}_{k+1}\nm x_{k}}-\eta}\cdot
\frac{\delta\eto{x_{k}\nm \wx_{k-1}}-\eta}
 {\raisebox{-5pt}{$\beta\eto{x_{k}\nm \wx_{k-1}}-\epsilon$}}=1,\quad
\end{equation}
which happens to encapsulate all the non-symmetric equations. For
instance:
\begin{itemize}
\item System (\ref{dRTL+ lq New introd}) appears from
(\ref{generalToda nonsymm}) with the choice $\eta=\infty$,
$\beta=(h-\alpha)/\alpha$ and $\epsilon=1/\alpha^2$. In the
quad-equations it is convenient to set $\delta=\alpha h$ and to
re-scale $\lambda\rightsquigarrow\alpha\lambda$.

\item System (\ref{dRTL+ m New introd}) appears from
(\ref{generalToda nonsymm}) with the choice $\beta=\eta=-1$,
$\delta=h$ and $\epsilon=1/\alpha$.

\item One gets from (\ref{generalToda nonsymm}) to the additive
equation (\ref{dRTL+ l New introd}) in two steps. On the first
step, one starts with the parameter values
$\beta=(h-\theta)/\theta$, $\eta=\theta/(\alpha-\theta)$,
$\delta=h\theta$, $\epsilon=1/(\alpha\theta)$, which leads to a
remarkable equation introduced in \cite{Su03}:
\begin{multline}\label{dRTL+ m2 New introd}
\displaystyle\frac{1+\theta h^{-1}\Big(\ueto{\wx_k\nm x_k}-1\Big)}
{1+\theta h^{-1}\Big(\ueto{x_k\nm\undertilde{x}_k}-1\Big)}=\nonumber\\
= \displaystyle\frac{1+\theta\alpha\eto{x_{k+1}\nm
x_k}}{1+\theta\alpha\eto{x_k\nm x_{k-1}}}\cdot
\frac{1+h(\theta-\alpha)\,\ueto{\undertilde{x}_{k+1}\nm
x_k}}{1+\alpha(\theta-h)\ueto{\undertilde{x}_{k+1}\nm x_k}}\cdot
\frac{1+\alpha(\theta-h)\eto{x_k\nm\wx_{k-1}}}
{\raisebox{-4pt}{$1+h(\theta-\alpha)\,\eto{x_{k}\nm\wx_{k-1}}$}}
\,.\tag{*}
\end{multline}
This equation interpolates between (\ref{dRTL+ lq New introd})
(corresponding to $\theta=\alpha$) and (\ref{dRTL+ l New introd})
(which corresponds to $\theta=0$). In the corresponding
quad-equations it is convenient to re-scale
$\lambda\rightsquigarrow\theta\lambda$. We remark that the last
equation is a time discretization of
\begin{eqnarray*}\label{RTL+ m2 New introd}
\ddot{x}_k & = & (1+\theta\dot{x}_k)\!\left(\!
(1+\alpha\dot{x}_{k+1})\,\displaystyle\frac{\eto{x_{k+1}\nm x_k}}
{\,\raisebox{-1mm}{$1+\theta\alpha\eto{x_{k+1}\nm x_k}$}}-
(1+\alpha\dot{x}_{k-1})\,\displaystyle\frac{\eto{x_k\nm x_{k-1}}}
{\,\raisebox{-1mm}{$1+\theta\alpha\eto{x_k\nm x_{k-1}}$}} \right.
 \nonumber\\
 & & \left. +\alpha(\theta-\alpha)\,
 \displaystyle\frac{\eto{2(x_{k+1}\nm x_k)}}
{\,\raisebox{-1mm}{$1+\theta\alpha\eto{x_{k+1}\nm x_k}$}}
-\alpha(\theta-\alpha)\,\displaystyle\frac{\eto{2(x_k\nm
x_{k-1})}} {\,\raisebox{-1mm}{$1+\theta\alpha\eto{x_k\nm
x_{k-1}}$}} \right),
\end{eqnarray*}
which in turn interpolates between the continuous time equations
(\ref{RTL+ lq New introd}) (for $\theta=\alpha$) and (\ref{RTL+ l
New introd}) (for $\theta=0$). On the second step, one performs in
equation (\ref{dRTL+ m2 New introd}) the limit $\theta\to 0$. In
this limit one should also re-scale the auxiliary variables
according to $\Psi\rightsquigarrow\lambda\Psi$,
$\Phi\rightsquigarrow\Phi/\lambda$, and $\lambda\rightsquigarrow
1+\theta\lambda/2$.

\item To transform (\ref{generalToda nonsymm}) to the rational
equation (\ref{dRTL+ dual New introd}) one makes the change of
variables
\[
x\leadsto \kappa x,
\]
accompanied by the change of parameters
\[
\beta\leadsto-1+\kappa\beta, \quad \delta\leadsto-1+\kappa\delta,
\quad \epsilon\leadsto-1+\kappa\epsilon,  \quad
\eta\leadsto-1+\kappa\eta.
\]
Sending $\kappa\to 0$, one arrives at
\[
\frac{\wx_{k}-x_{k}+\beta}{x_{k}-\undertilde{x}_{k}+\beta}\cdot
\frac{x_{k}-x_{k-1}+\epsilon}{x_{k+1}-x_{k}+\epsilon}\cdot
\frac{\undertilde{x}_{k+1}-x_{k}-\beta+\epsilon}
{\undertilde{x}_{k+1}-x_{k}-\delta+\eta}\cdot
\frac{x_{k}-\wx_{k-1}-\delta+\eta}{x_{k}-\wx_{k-1}-\beta+\epsilon}=1.
\]
Equation (\ref{dRTL+ dual New introd}) corresponds to the choice
\[
\beta=\eta=0, \quad \delta=-1/h, \quad \epsilon=1/\alpha.
\]
\end{itemize}

\section{Zero curvature representations}
\label{Sect zcr}

The construction of discrete Toda type systems on graphs from
systems of quad-equations allows one to find, in an algorithmic
way, discrete zero curvature representations for the former.
Indeed, each quad-equation can be viewed as a M\"obius
transformation of the field at one white vertex of a quad into the
field at the other white vertex, with the coefficients dependent
on the fields at the both black vertices. The $SL(2,\mathbb C)$
matrices representing these M\"obius transformations play then the
role of transition matrices across the edges connecting the black
vertices. The property (\ref{zero curv cond}) is satisfied
automatically, by construction.

Specializing this construction to the case of the regular
triangular lattice (see Figure \ref{fig:L3}), we denote by $L_k$
the transition matrix from $\Psi_k$ to $\Psi_{k+1}$, and by $V_k$
the transition matrix from $\Psi_k$ to $\widetilde{\Psi}_k$. In
this notation, the discrete zero representation reads:
\begin{equation}\label{eq: dzcr}
  \widetilde{L}_kV_k=V_{k+1}L_k,
\end{equation}
both parts representing the transition from $\Psi_k$ to
$\widetilde{\Psi}_{k+1}$ along two different paths. It is clear
that $L_k$ is the product of two matrices, the first corresponding
to the transition from $\Psi_k$ to $\Phi_{k+1}$ across the edge
$[x_k,\undertilde{x}_k]$, and the second corresponding to the
transition from $\Phi_{k+1}$ to $\Psi_{k+1}$ across the edge
$[x_k,\undertilde{x}_{k+1}]$, so that
\begin{equation}\label{eq: L gen}
L_k=L(x_k,\undertilde{x}_k,\undertilde{x}_{k+1};\lambda).
\end{equation}
Similarly, $V_k$ can be represented as the product of two
matrices, the first corresponding to the transition from $\Psi_k$
to $\widetilde{\Phi}_{k}$ across the edge $[x_k,x_{k-1}]$, and the
second corresponding to the transition from $\widetilde{\Phi}_{k}$
to $\widetilde{\Psi}_{k}$ across the edge $[x_k,\wx_{k-1}]$, so
that
\begin{equation}\label{eq: V gen}
V_k=V(x_k,x_{k-1},\wx_{k-1};\lambda).
\end{equation}
The matrices $L_k$, $V_k$ for a given discrete relativistic Toda
type equation can be computed in a straightforward way, as soon as
the generating system of quad-equations is known. Theorem \ref{Th
main} provides us with the means for this goal.

The resulting zero curvature representations possess an additional
remarkable property. It is well known (see, e.g., \cite{ABS03},
\cite{AS04}) that the discrete relativistic Toda type equations
possess a Lagrangian (variational) interpretation with a discrete
Lagrange function
\begin{equation}
\mathcal L(x,\undertilde{x})=\sum_{k\in\mathbb
Z}\Big(\Lambda_1(x_k-\undertilde{x}_k)
-\Lambda_2(\undertilde{x}_{k+1}-\undertilde{x}_k)
-\Lambda_3(\undertilde{x}_{k+1}-x_k)\Big).
\end{equation}
Here $\Lambda_1$, $\Lambda_2$, $\Lambda_3$ are antiderivatives of
the functions $F$, $G$, $H$ in the general equation (\ref{dRTL gen
intro}). The canonically conjugate momenta and the Lagrangian form
of equations of motion are given by
\[
p_k=\frac{\partial\mathcal L(x,\undertilde{x})}{\partial x_k}=
-\frac{\partial\mathcal L(\wx,x)}{\partial x_k},
\]
which specializes in our case to
\begin{eqnarray}
p_k & = & F(x_k-\undertilde{x}_k)+ H(\undertilde{x}_{k+1}-x_k)
\label{eq: p can 1}\\
    & = &
    F(\wx_k-x_k)+H(x_k-\wx_{k-1})-G(x_{k+1}-x_k)+G(x_k-x_{k-1}).
\label{eq: p can 2}
\end{eqnarray}
\begin{theorem}
For all discrete relativistic Toda type systems, the transition
matrix $L_k$ from equation (\ref{eq: L gen}) is local, when
expressed in terms of canonically conjugate variables:
\begin{equation}
L_k=L(x_k,p_k;\lambda).
\end{equation}
Moreover, as a matter of fact, the matrix $L_k$ does not depend on
the time discretization parameter $h$, so that the corresponding
Lagrangian maps $(x,p)\mapsto(\wx,\widetilde{p})$ belong to the
same integrable hierarchies as their respective continuous time
Hamiltonian counterparts. In other words, these Lagrangian maps
serve as B\"acklund transformations for the respective Hamiltonian
flows, the B\"acklund parameter being the time step $h$.
\end{theorem}
{\bf Proof} of this theorem is obtained again via direct
computations on the case-by-case basis. For all cases where the
local discrete zero curvature representation was known (those
cases are listed in \cite{Su03}) the results obtained from the
system of quad-equations coincide with the previously available
ones. Therefore, we illustrate the claims of the theorem with the
case where the local discrete zero curvature representation was
not known previously, namely with the rational systems (\ref{dRTL+
dual New introd}), (\ref{dRTL- dual New introd}). It is useful to
keep in mind that these two systems come as two elementary flows
(a positive and a negative ones) of the same hierarchy
\cite{Su03}.\hfill $\Box$
\medskip

{\bf Equation (\ref{dRTL+ dual New introd}).} \quad The Lagrangian
form reads:
\begin{eqnarray}
h\eto{p_{k}} & = & (x_{k}-\undertilde{x}_{k})\cdot
\frac{1+h(\undertilde{x}_{k+1}-x_k)}{1+\alpha(\undertilde{x}_{k+1}-x_k)}
\label{dRTL+ dual p1}\\
  & = & (\wx_{k}-x_{k})\cdot
\frac{1+\alpha(x_{k}-x_{k-1})}{1+\alpha(x_{k+1}-x_{k})}\cdot
\frac{1+h(x_{k}-\wx_{k-1})} {1+\alpha(x_{k}-\wx_{k-1})}\,.
\label{dRTL+ dual p2}
\end{eqnarray}
The transition matrices of the zero curvature representation of
this map read:
\begin{equation}\label{dRTL+ dual L}
L_{k}=\begin{pmatrix}-\lambda+x_{k}& & &\lambda^2+\lambda\alpha
\eto{p_{k}}-
(x_{k}-\alpha\eto{p_{k}})x_{k}-\eto{p_{k}}\\
1 & & &-\lambda-x_{k}+\alpha \eto{p_{k}}\end{pmatrix}
\end{equation}
and
\begin{equation}\label{dRTL+ dual V}
V_{k}=I+h\begin{pmatrix}-\lambda+x_{k}& & &
\lambda^2+\lambda(x_{k}-\wx_{k-1}+\alpha\eto{\widetilde{p}_{k-1}})
-(\wx_{k-1}-\alpha\eto{\widetilde{p}_{k-1}})x_{k}\\
1& & &
-\lambda-\wx_{k-1}+\alpha\eto{\widetilde{p}_{k-1}}\end{pmatrix}.
\end{equation}
Note that in the limit $h\to 0$ one obtains a zero curvature
representation (\ref{zcr intro}) for the Hamiltonian form of
equation (\ref{RTL+ dual New introd}) with the same matrix $L_k$
as in (\ref{dRTL+ dual L}) and with the matrix
\begin{equation}\label{RTL+ dual M}
M_{k}=\begin{pmatrix}-\lambda+x_{k}& & &
\lambda^2+\lambda(x_{k}-x_{k-1}+\alpha\eto{{p}_{k-1}})
-(x_{k-1}-\alpha\eto{{p}_{k-1}})x_{k}\\
1& & & -\lambda-x_{k-1}+\alpha\eto{{p}_{k-1}}\end{pmatrix}.
\end{equation}
To the best of our knowledge, this result (and even its
non-relativistic particular case $\alpha=0$) was previously
unknown.
\medskip

{\bf Equation (\ref{dRTL- dual New introd}).} \quad The Lagrangian
form reads:
\begin{align}
h\eto{p_{k}}&=\frac{\wx_{k}-x_{k}}
{1+\alpha(\alpha+h)h^{-1}(\wx_{k}-x_{k})}\cdot
(1+(\alpha+h)(x_{k}-\wx_{k-1}))
\label{dRTL- dual p1}\\
&=\frac{x_{k}-\undertilde{x}_{k}}
{1+\alpha(\alpha+h)h^{-1}(x_{k}-\undertilde{x}_{k})}\cdot
\frac{1+\alpha({x}_{k+1}-{x}_{k})}{1+\alpha({x}_{k}-{x}_{k-1})}\cdot
(1+(\alpha+h)(\undertilde{x}_{k+1}-{x}_{k})).
 \label{dRTL- dual p2}
\end{align}
This Lagrangian map admits a discrete zero curvature
representation
\begin{equation} \label{RTL-RZCR}
W_{k+1}\widetilde{L}_{k}=L_{k}W_{k}
\end{equation}
with the same transition matrix $L_k$ as in (\ref{dRTL+ dual L})
and with
\begin{multline}
W_{k}=I-\frac{h}{(1-2\alpha\lambda)\Big(1+(\alpha+h)(x_{k}-\wx_{k-1}-
\alpha\eto{p_{k}})\Big)}\\
\quad\times \begin{pmatrix}\lambda+x_{k}- \alpha \eto{p_{k}} & & &
\lambda^2+\lambda(x_{k}-\wx_{k-1}-\alpha
\eto{p_{k}})-(x_{k}-\alpha
\eto{p_{k}})x_{k-1}\\
1 & & & \lambda-\wx_{k-1}\end{pmatrix}.
\end{multline}
Again, in the limit $h\to 0$ one obtains a zero curvature
representation (\ref{zcr intro}) for the Hamiltonian form of
equation (\ref{RTL- dual New introd}) with the same matrix $L_k$
as in (\ref{dRTL+ dual L}) and with the matrix
\begin{multline}
M_{k}=\frac{1}{(1-2\alpha\lambda)\Big(1+\alpha(x_{k}-x_{k-1}-
\alpha\eto{p_{k}})\Big)}\\
\quad\times \begin{pmatrix}\lambda+x_{k}-\alpha \eto{p_{k}} & & &
\lambda^2+\lambda(x_{k}-x_{k-1}-\alpha \eto{p_{k}})-(x_{k}-\alpha
\eto{p_{k}})x_{k-1}\\
1 & & & \lambda-x_{k-1}\end{pmatrix}.
\end{multline}
Also these results seem to be previously unknown, even in the
continuous time case.

\section{Conclusions}
\label{Sect concl}

In this paper, we clarified the origin of all non-symmetric
discrete equations of the relativistic Toda type from 3D
consistent systems of quad-equations. Unlike the symmetric case,
the three coordinate planes carry here three different
quad-equations, so that a more general understanding of the 3D
consistency than usual is required. Note that this more general
concept has already been discussed and laid into the basis of a
classification procedure in \cite{ABS09}, see also examples
discussed in \cite{At08}. However, the classification only has
been performed for the so called systems of type Q in \cite{ABS09}
(all edge biquadratics non-degenerate, see the original paper for
details). Examples which arose in the present work demonstrate
that also the systems of type H (with some of the edge
biquadratics being degenerate) are of a considerable importance,
which calls for a complete classification of this case, as well.
We plan to turn to this tedious task in our future work.

\label{lastpage}


\begin{thebibliography}{10}
\markboth{R. Boll and Yu.B. Suris}{Discrete Toda systems}

\bibitem{A99} V.E.~Adler, {\em Legendre transformations on the triangular
lattice}, Funct. Anal. Appl., {\bf 34} (1999), pp. 1--9.

\bibitem{A00} ---------,  {\em On the structure of the B\"acklund
transformations for the relativistic lattices},  J. Nonlin. Math.
Phys. 7 (2000), pp. 34--56.

\bibitem{A01} ---------,  {\em Discrete equations on planar graphs},
J. Phys. A: Math. Gen. 34 (2001), pp. 10453--10460.

\bibitem{ABS03} V.E.~Adler, A.I.~Bobenko, Yu.B.~Suris,
{\em Classification of integrable equations on quad-graphs.
 The consistency approach}, Comm. Math. Phys. 233 (2003), pp. 513--543.

\bibitem{ABS09} V.E.~Adler, A.I.~Bobenko, Yu.B.~Suris,
{\em Discrete nonlinear hyperbolic equations. Classification of
integrable cases}, Funct. Anal. Appl. 43 (2009), pp. 3--17.

\bibitem{ASh97} V.E.~Adler and A.B.~Shabat, {\em Generalized Legendre
transformations}, Theor. Math. Phys. 112 (1997), pp. 935--948.

\bibitem{AS04} V.E.~Adler, Yu.B.~Suris, {\em Q4: integrable master equation
related to an elliptic curve}, Intern. Math. Research Notices, Nr.
47 (2004), pp. 2523--2553.

\bibitem{At08} J.~Atkinson, {\em B\"acklund transformations for
integrable lattice equations}, J. Phys. A: Math. Theor. 41 (2008)
135202, 8 pp.

\bibitem{BMS05} A.I.~Bobenko, Ch.~Mercat, Yu.B.~Suris, {\em Linear and
nonlinear theories of discrete analytic functions. Integrable
structure and isomonodromic Green's function},  J. Reine Angew.
Math.,  583 (2005), pp. 117--161.

\bibitem{BS02} A.I.~Bobenko, Yu.B.~Suris. {\em Integrable systems on
quad-graphs}, Intern. Math. Research Notices, Nr. 11 (2002), pp.
573--611.

\bibitem{BS08} ---------, {\em Discrete Differential Geometry.
Integrable Structure}, Graduate Studies in Mathematics , Vol. 98.
AMS, Providence, 2008.

\bibitem{B08} R.~Boll, {\em Embedding of Non-Symmetric Discrete
Toda Systems in Multidimensional Lattices}, Diploma thesis, TU
M\"unchen, 2009.

\bibitem{N02} F.~Nijhoff, {\em Lax pair for the Adler (lattice
Krichever--Novikov) system}, Phys. Lett. A 297 (2002), pp. 49--58.

\bibitem{Ru90} S.N.M.~Ruijsenaars, {\em Relativistic Toda
systems}, Commun. Math. Phys. 133 (1990), pp. 217--247.

\bibitem{Su96} Yu.B.~Suris, {\em A discrete-time relativistic Toda
lattice}, J. Phys. A: Math. and Gen. 29 (1996), pp. 451--465.

\bibitem{Su97} ---------, {\em New integrable systems related to the
relativistic Toda lattice}, J. Phys. A: Math. and Gen. 30 (1997),
pp. 1745--1761.

\bibitem{Su99} ---------, {\em $R$-matrix hierarchies, integrable
lattice systems, and their integrable discretizations}, in {\it
Symmetries and Integrability of Difference Equations}, P.Clarkson
and F.Nijhoff, eds, Cambridge Univ. Press, 1999, pp. 79--94.

\bibitem{Su03} ---------, {\em The Problem of Integrable
Discretization. Hamiltonian Approach}, Progress in Mathematics
Vol. 219, Birkh\"auser, Basel, 2003.

\end{thebibliography}
\end{document}